\documentclass[twocolumn]{aastex631} %linenumbers
\hypersetup{linkcolor=blue,citecolor=blue,filecolor=blue,urlcolor=blue}
\pdfoutput=1
\newcommand{\angstrom}{\textup{\AA}}
\newcommand{\hii}{H\,{\sc ii}\rm}
\newcommand{\hi}{H\,{\sc i}\rm}

\usepackage{graphicx}
\usepackage{amsmath}  %%% you should have this, very basic math package
\usepackage{xcolor}    %%% for coloring my text parts
\usepackage{subfigure}
\usepackage{orcidlink}
\usepackage{CJK}
\shortauthors{Sextl et al.}
\shorttitle{TYPHOON: M83}
\begin{document}
\begin{CJK*}{UTF8}{gbsn}
%\title{The TYPHOON Stellar Population Synthesis Survey. II. The Nearby Grand Design Barred Spiral M83 - Central and Azimuthal Inhomogeneities of the Young Stellar Population}

\title{The TYPHOON Stellar Population Synthesis Survey. II. Pushing Full Spectral Fitting to the Limit in the Nearby Grand Design Barred Spiral M83}
\accepted{26 May 2025}
\correspondingauthor{Eva Sextl}
\email{sextl@usm.lmu.de}

\author{Eva Sextl \orcidlink{0009-0001-5618-4326}}
\affiliation{Universit\"ats-Sternwarte, Fakult\"at f\"ur Physik, Ludwig-Maximilians Universit\"at M\"unchen, Scheinerstr. 1, 81679 M\"unchen, Germany}
\author{Rolf-Peter Kudritzki}
\affiliation{Universit\"ats-Sternwarte, Fakult\"at f\"ur Physik, Ludwig-Maximilians Universit\"at M\"unchen, Scheinerstr. 1, 81679 M\"unchen, Germany}
\affiliation{Institute for Astronomy, University of Hawaii at Manoa, 2680 Woodlawn Drive, Honolulu, HI 96822, USA}
\author{Fabio Bresolin \orcidlink{0000-0002-5068-9833}}
\affiliation{Institute for Astronomy, University of Hawaii at Manoa, 2680 Woodlawn Drive, Honolulu, HI 96822, USA}
\author{Kathryn Grasha \orcidlink{0000-0002-3247-5321}}
\affiliation{Research School of Astronomy and Astrophysics, Australian National University, Canberra, ACT 2611, Australia}
\affiliation{ARC Centre of Excellence for All Sky Astrophysics in 3 Dimensions (ASTRO 3D), Australia}
\affiliation{Visiting Fellow, Harvard-Smithsonian Center for Astrophysics, 60 Garden Street, Cambridge, MA 02138, USA}
\author{Hye-Jin Park \orcidlink{0000-0002-9809-6631}}
\affiliation{Research School of Astronomy and Astrophysics, Australian National University, Canberra, ACT 2611, Australia}
\affiliation{ARC Centre of Excellence for All Sky Astrophysics in 3 Dimensions (ASTRO 3D), Australia}
\author{Qian-Hui Chen (陈千惠) \orcidlink{0000-0002-4382-1090}}
\affiliation{Research School of Astronomy and Astrophysics, Australian National University, Canberra, ACT 2611, Australia}
\affiliation{ARC Centre of Excellence for All Sky Astrophysics in 3 Dimensions (ASTRO 3D), Australia}
\author{Andrew J. Battisti \orcidlink{0000-0003-4569-2285}}
\affiliation{Research School of Astronomy and Astrophysics, Australian National University, Canberra, ACT 2611, Australia}
\affiliation{International Centre for Radio Astronomy Research
(ICRAR), University of Western Australia, M468, Crawley, WA, Australia}
\affiliation{ARC Centre of Excellence for All Sky Astrophysics in 3 Dimensions (ASTRO 3D), Australia}
\author{Mark Seibert \orcidlink{0000-0002-1143-5515}}
\affiliation{The Observatories, Carnegie Institution for Science, 813 Santa Barbara Street, Pasadena, CA 91106, USA}
\author{Barry F. Madore \orcidlink{0000-0002-1576-1676}}
\affiliation{The Observatories, Carnegie Institution for Science, 813 Santa Barbara Street, Pasadena, CA 91106, USA}
\author{Jeffrey A. Rich \orcidlink{0000-0002-5807-5078}}
\affiliation{The Observatories, Carnegie Institution for Science, 813 Santa Barbara Street, Pasadena, CA 91106, USA}
%\author{Rachael L. Beaton \orcidlink{0000-0002-1691-8217}}
%\affiliation{Department of Astrophysical Sciences, 4 Ivy Lane, Princeton University, Princeton, NJ 08544, USA}
%\affiliation{The Observatories, Carnegie Institution for Science, 813 Santa Barbara Street, Pasadena, CA 91106, USA}

%%%%%%%%%%%%%%%%%%%%%%%%%%%%%%%%%%%%%%%%%%%%%%%%%%%%%%%%%%%%%%%%%%%%%%%%%%%%

\begin{abstract}
We apply population synthesis techniques to analyze TYPHOON long slit spectra of the starburst barred spiral galaxy M83. The analysis covers a central square of 5 arcmin side length. We determine the spatial distribution of dust through the analysis of reddening and extinction, together with star formation rates, ages, and metallicities of young and old stellar populations. For the first time, a spatial one-to-one comparison of metallicities derived from full-spectral fitting techniques with those obtained from individual young stellar probes has been carried out. The comparison with blue supergiant stars, young massive star clusters, and super star clusters shows a high degree of concordance when wavelength coverage in the $B$-band is available. The metallicity of the young population is supersolar and does not show a radial metallicity gradient along the investigated part of the disk, in agreement with our chemical evolution model. However, a notable decrease in metallicity is observed in a tightly confined region at the galaxy center, coinciding with circumnuclear orbits. We attribute this to matter infall either from the circumgalactic medium or a dwarf galaxy interloper or, alternatively, to AGN-interrupted chemical evolution. We confirm the presence of a dust cavity with a diameter of 260~pc close to the galaxy center. Dust absorption and molecular CO emission are spatially well correlated. We find an anticorrelation between R$_V$, the ratio of dust attenuation to reddening, and the emission strength of molecular species present in photo-dissociation regions. We confirm our results by using alternative fitting algorithms and stellar libraries.
\end{abstract}



%%%%%%%%%%%%%%%%%%%%%%%%%%%%%%%%%%%%%%%%%%%%%%%%%%%%%%%%%%%%%%%%%%%%%%%%%%%%

\section{Introduction} \label{sec:intro}

The spatial distribution of gaseous and stellar metallicity provides unique information on the evolution of galaxies. For instance, negative metallicity gradients in the disks of massive star-forming galaxies, i.e. the linear decline of the logarithm of metallicity with galactocentric distance, are commonly used to constrain galaxy growth rates, star formation efficiency, and the effects of galactic winds and matter infall (see, for example, \citealt{Kang2021,Kang2023,Kudritzki2015, Molla2017}). Although it seems that these gradients are universal when normalized to a characteristic scale length \citep{SanchezS2014,Ho2015}, there has been growing evidence that the concept of negative gradients is too simple. Extensive studies of \hii\ region metal abundances have shown clear deviations from single gradients and revealed a flattening of the spatial metallicity distribution in the central and also in the outer parts of galaxy disks \citep{Bresolin2009b,Bresolin2012,Goddard2011, SanchezM2016,SanchezM2018, Franchetto2021, groves2023, Chen2023}. Positive metallicity gradients were encountered at higher redshifts and may indicate the influence of galaxy interaction and very strong infall \citep{Tissera2022,Vallini2024}. In addition,  deviations from the azimuthal homogeneity of metallicity, sometimes associated with spiral arms, have been observed \citep{Ho2017,Ho2018,Kreckel2019,SanchezM2020}, indicating the importance of localized enrichment through strong star formation activity or metal dilution through matter infall. Theoretical chemical evolution models indicate that spiral density waves may indeed introduce metallicity inhomogeneities \citep{Ho2017,Spitoni2019,Molla2019}.

Given the relevance of these intriguing results, it is important to carry out an independent investigation using metallicity information obtained from the quantitative spectroscopy of the young stellar population in galaxies. Indeed, spectroscopy of individual blue and red supergiant stars has resulted in accurate measurements of massive star metallicities and metallicity gradients and revealed that \hii\ region abundance measurements are frequently affected by systematic uncertainties \citep{Kudritzki2008,Kudritzki2012,Kudritzki2016,Bresolin2009a,Bresolin2016,Bresolin2022,Gazak2015,Davies2017, Lardo2015,Liu2022}. At the same time, these observations did not allow one to investigate deviations from azimuthal homogeneity because of the small number of individual stellar targets distributed over the galaxy surface. 

A very promising alternative is the use of integral field units (IFU) and the population synthesis analysis of the spectra of the integrated stellar populations. To be precise, we apply techniques for full-spectral fitting. It compares combinations of single-burst stellar population models (SSPs) to observed spectra in order to extract physical properties of the stellar populations, such as age, metallicity, and star formation history, either for entire galaxies or spatially resolved regions. The power of this method has been demonstrated in a large number of publications (see, for instance, \citealt{bittner2020}, \citealt{Carrillo2020}, \citealt{parikh2021}, \citealt{Emsellem2022}, \citealt{Pessa2023}, \citealt{westmoquette2011}, \citealt{sanchez2014}, \citealt{Thaina2023}), which provided unique information about the distribution of ages and metallicities averaged over all populations, old and young, in the galaxies investigated. However, for our purpose, which focuses on the properties of the young stellar population, it is important to use observations and a methodology that are capable of disentangling the contributions of young and old stars over a large spatial area of the galaxies. 

The TYPHOON survey (see Section \ref{sec:obs} for more details), which uses stepwise combined long-slit spectra of galaxies, is ideal for this purpose. It is based on observations covering  a large field of view across a wide wavelength range, including the blue to near-UV parts of the stellar spectral energy distributions (SEDs), which are absolutely crucial to constrain the properties of the young stars from the spectra of the integrated populations.
We started a project to systematically investigate the disentangled properties of the young and old stellar populations in nearby star-forming galaxies. In a first step, we have studied the barred spiral NGC~1365 and have obtained crucial information about the distribution of dust, star formation rates, metallicities, and ages (\citealt{Sextl2024}, hereafter Paper I). Most intriguingly, we found a significant drop in the metallicity of the young population in a closely confined region in the center of the galaxy, which we interpreted as the result of infall of metal-poor gas along the bar or AGN interrupted chemical evolution, or both. Now in the next step, we turn to M83. 

M83 (=\, NGC 5236) is the nearest face-on, grand-design barred spiral galaxy showing evidence of intense recent star formation activity and ongoing star formation bursts in its center and two spiral arms over at least the last 10 Myr \citep{Dopita2010}.  A schematic illustration of M83's most prominent (gas) structures can be found in \citet{Koda2023}.

This galaxy has experienced six supernovae in the last century as a result of its high star-forming activity. Detailed optical spectroscopy of blue supergiant stars (BSGs, \citealt{Bresolin2016}), together with the study of UV and optical integrated spectra of young massive clusters (YMCs, \citealt{Hernandez2018,Hernandez2019}), the NIR spectroscopic analysis of the red supergiant-dominated super star clusters (SSCs, \citealt{Davies2017}), and the exploration of metal emission and absorption lines from the interstellar medium (ISM, \citealt{Bresolin2005,Bresolin2016,Hernandez2021,Grasha2022}) indicate high metallicity, clearly higher than the metallicity in the sun and the young stars and the ISM of the solar neighborhood \citet{Asplund2009}. However, these investigations disagree with regard to the metallicity gradient in the inner part of the galactic disk, located within two-thirds of the isophotal radius. The SSCs indicate a flat metallicity distribution without any gradient, the ISM emission and absorption lines point to a very weak gradient, whereas the YMCs and BSGs provide evidence for a stronger gradient. The low number of objects and their sparse distribution is possibly affecting the conclusions about the metallicity distribution. A more complete spatial coverage of the young stellar population over the galaxy might provide a clearer picture.
In addition, comparing metallicities obtained from individual stellar sources (BSGs, YMCs, SSCs) and cospatial population synthesis will allow for a crucial test of the accuracy of our population synthesis technique. Other substantial tests on full-stellar fitting itself are difficult and often limited to mock observations \citep{Wilkinson2015} or simpler systems of globular clusters \citep{Koleva2008, Cezario2013, Geraldo2020}. In a rare opportunity, we will compare the metallicities of the young stellar population (age $\le$ 100 Myr) with those of the stellar probes at their specific positions in the galaxy. For this, a substantial part of M83 has to be observed in an IFU-like manner with a large spectral coverage, especially in the blue part of the optical spectrum. This approach will also enable us to investigate the properties and distribution of dust, as well as to look for potential correlations with the ISM molecular gas. In addition, we will also study the star formation properties of our target galaxy.

We describe the observations in Section \ref{sec:obs}. Here we also summarize the global properties of M83 and give a brief overview of the population synthesis analysis technique that we applied. The main results are presented in two sections. Section \ref{sec:results} analyzes the complex galactic distribution of interstellar dust, star formation rates and stellar ages, making a comparison with the distribution of molecular and cold atomic gas. Section~\ref{sec:results_2} discusses the spatial distribution of stellar metallicity and compares with that inferred from the spectroscopy of individual supergiant stars and stellar clusters. In addition, Section \ref{sec:approaches} provides a test of the population synthesis method by comparing with an alternative fitting algorithm and single stellar population spectra, and Section \ref{sec:lightweightmetal} illustrates the potential problems deriving from the use of simple luminosity-weighted averages of stellar metallicity. Section \ref{sec:wcoverage} investigates the importance of the wavelength range of the analyzed stellar spectra, and, finally, Section \ref{sec:discussion} concludes the paper with a summary and discussion. 

% - - - - - - - - - - - - - - - - - - - - - - - - - - - - - - - - - - - - - - - - - 
\begin{figure}[htb!]
\medskip
	\center \includegraphics[width=0.95\columnwidth]{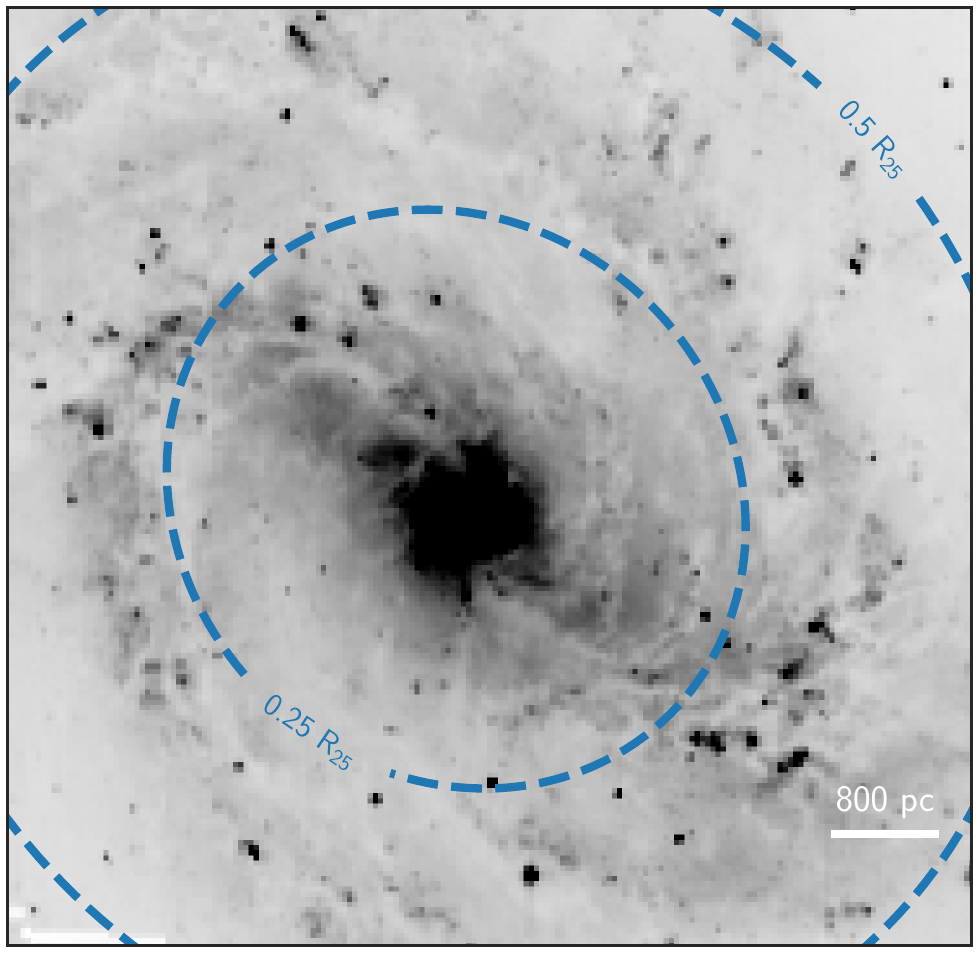}\medskip
    \center \includegraphics[width=0.95\columnwidth]{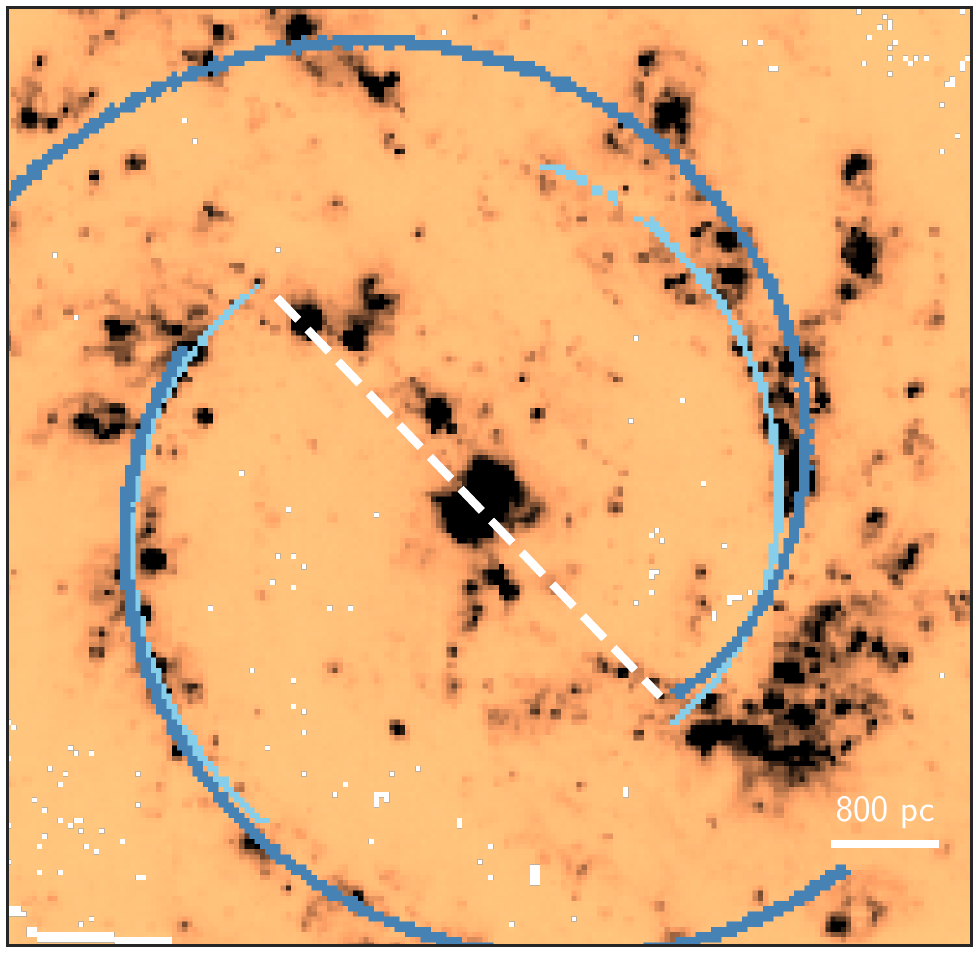}\medskip
	\caption{$V$-band image (top) and H${\alpha}$ line flux (bottom) of the inner disk of M83 created from the TYPHOON data cube. Top: The two dashed ellipses indicate galactocentric distances of 2.25 and 4.51 kpc, corresponding to isophotal radii of 0.25 and 0.50 R$_{25}$, respectively. Bottom: Two spiral arms using H${\alpha}$ are fitted with a logarithmic spiral (light-blue). In addition, similar fits of ALMA - CO(2-1) observations are over-plotted in dark blue. A schematic presentation of the bar is shown as white dashed line.
    East is towards the left and north towards the top in both panels.} \label{fig:Vband}
\end{figure}
% - - - - - - - - - - - - - - - - - - - - - - - - - - - - - - - - - - - - - - - - - 

%\begin{figure}[htb!]
%\medskip
%    \center \includegraphics[width=0.95\columnwidth]{figure2_halpha.pdf}\medskip
%	\caption{H${\alpha}$ image of the inner disk of M83 obtained from the TYPHOON data. Two spiral arms  are fitted with a logarithmic spiral (light-blue). In addition, similar fits of ALMA - CO(2-1) observations are over-plotted in dark blue. Orientation as in previous figure. } \label{fig:Halpha} 
%\end{figure}
 
% - - - - - - - - - - - - - - - - - - - - - - - - - - - - - - - - - - - - - - - - - 
\begin{figure}[htb!]
 \smallskip
	\center \includegraphics[width=0.98\columnwidth]{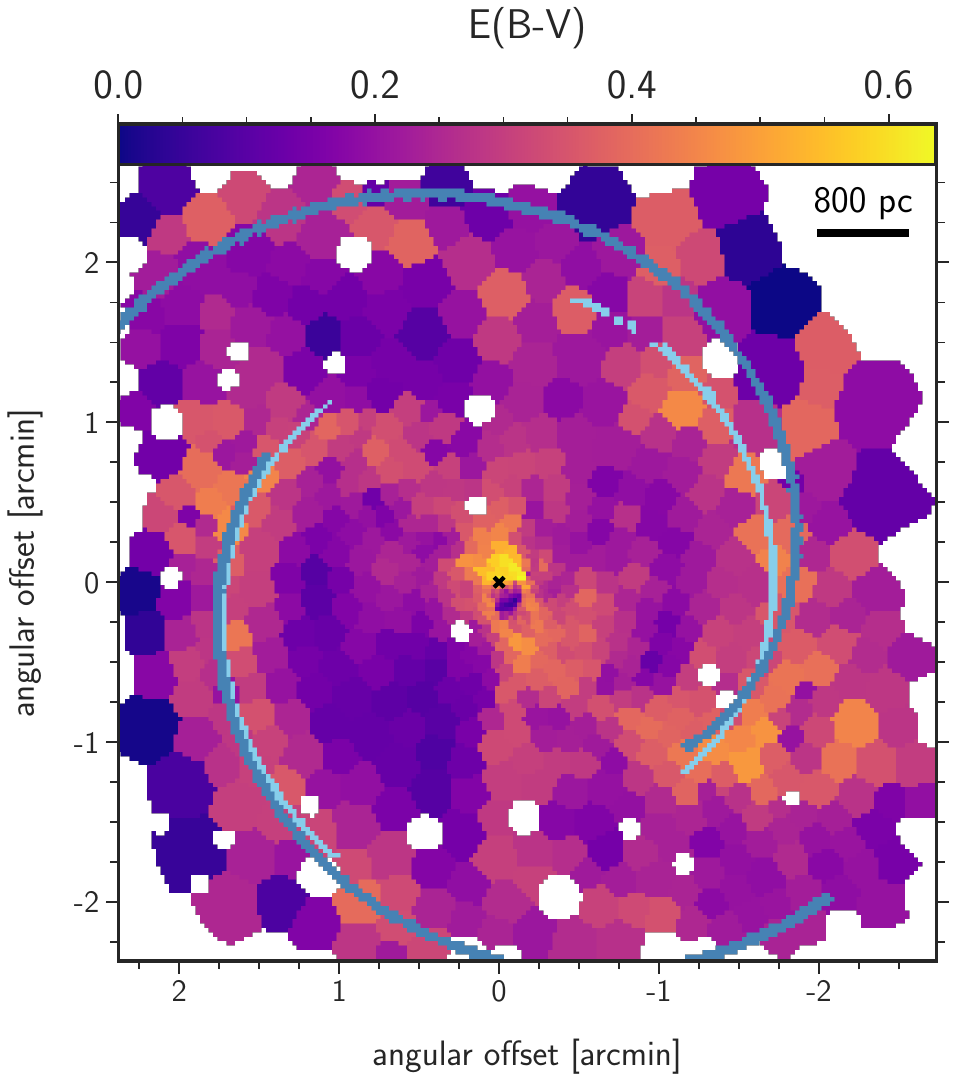}\medskip
	\caption{E(B-V) reddening map of M83 obtained from our TYPHOON population synthesis fit. The galaxy center defined by the peak of visible continuum surface brightness is at RA(J2000)\,=\,13$^h\,$37$^m\,$00\fs95 and DEC(J2000)\,=\,$-$29$^o$\,51'\,55\farcs50 \citep{Diaz2006} and indicated by a black cross. The spiral arms of Figure \ref{fig:Vband} are also indicated. Some areas are masked out as they are dominated by strong point sources and would therefore potentially lead to problematic fits (see text).}  \label{fig:redden1}
\end{figure}

\section{Global Properties of M83, Observations and Analysis Method} \label{sec:obs}

M83 is a moderately massive galaxy with a stellar mass of $\log( M_*/ M_\odot)$ = $10.41 \pm 0.10$ (\citealt{Leroy2019}; see also \citealt{Bresolin2022}), an isophotal radius of $6.44 \pm 0.05$ arcmin (\citealt{deVaucouleurs1991}, 9.01~kpc at the adopted distance), an inclination angle of 24 degrees and a position angle of $45 \pm 5$ degrees \citep{Comte1981}. \citet{Callanan2021} report, based on a review of existing data, a mean SFR of $0.8 \, M_{\odot} \, \textrm{yr}^{-1}$ for the central $\sim 500$ pc of M83. The distance to M83 of d = $4.81 \pm 0.2$ Mpc has been determined from blue supergiant stars (\citealt{Bresolin2016}, with the revision in \citealt{Sextl2021} and \citealt{Bresolin2022}). This value is in agreement with those derived by other methods, such as the tip of the red giant branch (EDD data base, \citealt{Tully2009}) or the Planetary Nebulae luminosity function \citep{Herrmann2008}. 

\subsection{Observations} \label{subsec:obs_Typhoon}

The TYPHOON survey observations (P.I. B. Madore) utilize the Las Campanas du Pont 2.5m telescope and the Wide Field CCD imaging spectrograph with a custom slit of 18 arcmin in length and 1.65 arcsec in width on the sky. To construct 3D data cubes with 1.65$\times$1.65 arcsec$^2$ spaxels the slit scans progressively across the galaxies (Progressive Integral Step Method, PrISM) with the slit orientation typically in the north-south direction. At the distance of M83, 1.65 arcsec is equivalent to 38.5 pc. The final datacubes contain spectra in the wavelength range of $3650$ to $8150$ \AA.  Their spectral resolution is relatively low and corresponds to a FWHM of 8.2 \AA. It does not allow us to constrain the line-broadening effects of stellar velocity dispersion. More details of the observational method are described in \cite{Grasha2023} and \cite{Chen2023}. 

Figure \ref{fig:Vband} (top) provides a $V$-band continuum image constructed from parts of the TYPHOON data cube. We restrict our study to this high surface brightness part of the disk, within $\sim$4~kpc from the center, because we need a minimum signal-to-noise ratio of our spectra for the analysis, as specified below. 
Additionally, the H${\alpha}$ line flux measured from each spaxel is shown in Figure \ref{fig:Vband} (bottom). The darker the pixel, the higher the  line flux obtained with the LZIFU fitting tool \citep{Ho2016}. The values saturate at $1.2 \cdot 10^{-14}$ erg/s/cm$^2$ for better visibility of the morphological structures. M83 is a barred spiral galaxy. In visible wavelengths, including H${\alpha}$, the spiral arms and clumpy star-forming regions dominate the view, making the bar less apparent. However, when observed in the infrared, the bar stands out clearly as a distinct structure of older stars and dust, cutting through the center of the galaxy \citep{Frick2016, Gallais1991}. \\
It should be emphasized that the transition regions, extending from the bar toward the beginning of the spiral arm, host prominent complexes of \hii\ regions. We note that the ionizing flux in the northeast edge of the bar is slightly less pronounced than in the equivalent area in the southwest. However, ALMA observations of both areas examined in \citet{Koda2023} revealed that these filamentary structures harbor similarly huge amounts of molecular gas and are probably caused by the overall galactic dynamics.

For our spectral analysis we use the spectral range from 4000~\AA\ to 7070~\AA. Here, the flux-calibrated spectra of the data cube have the highest signal-to-noise ratio. We stress that compared to the range 4800~\AA\ to 7000~\AA, typically used in studies carried out with the extremely powerful MUSE IFU attached to the ESO VLT, this is a crucial extension towards the blue which enables a more accurate characterization of the young stellar population. We will investigate this in more detail in Section \ref{sec:wcoverage}.

% - - - - - - - - - - - - - - - - - - - - - - - - - - - - - - - - - - - - - - - - - 
\begin{figure}[htb!]
       \medskip
	\center \includegraphics[width=0.93\columnwidth]{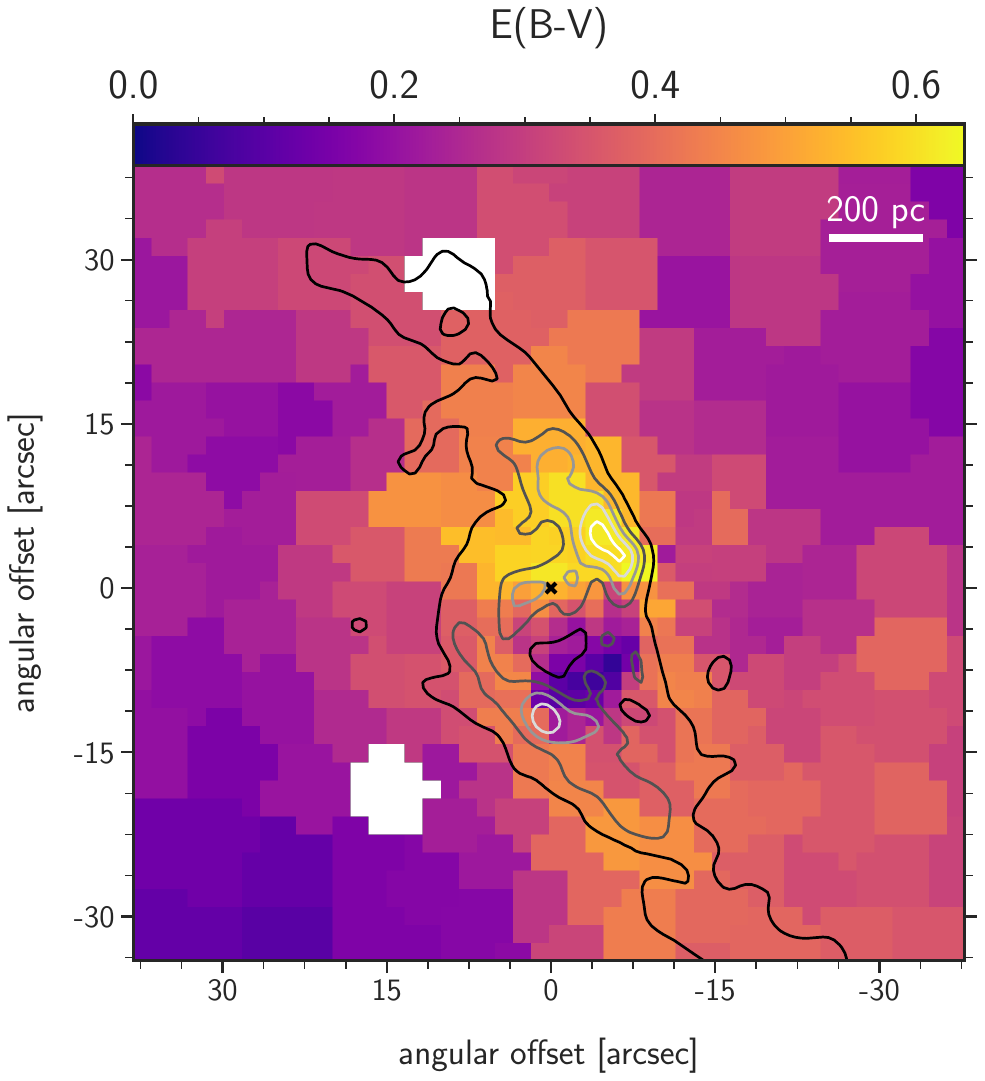}\medskip
    \center \includegraphics[width=0.93\columnwidth]{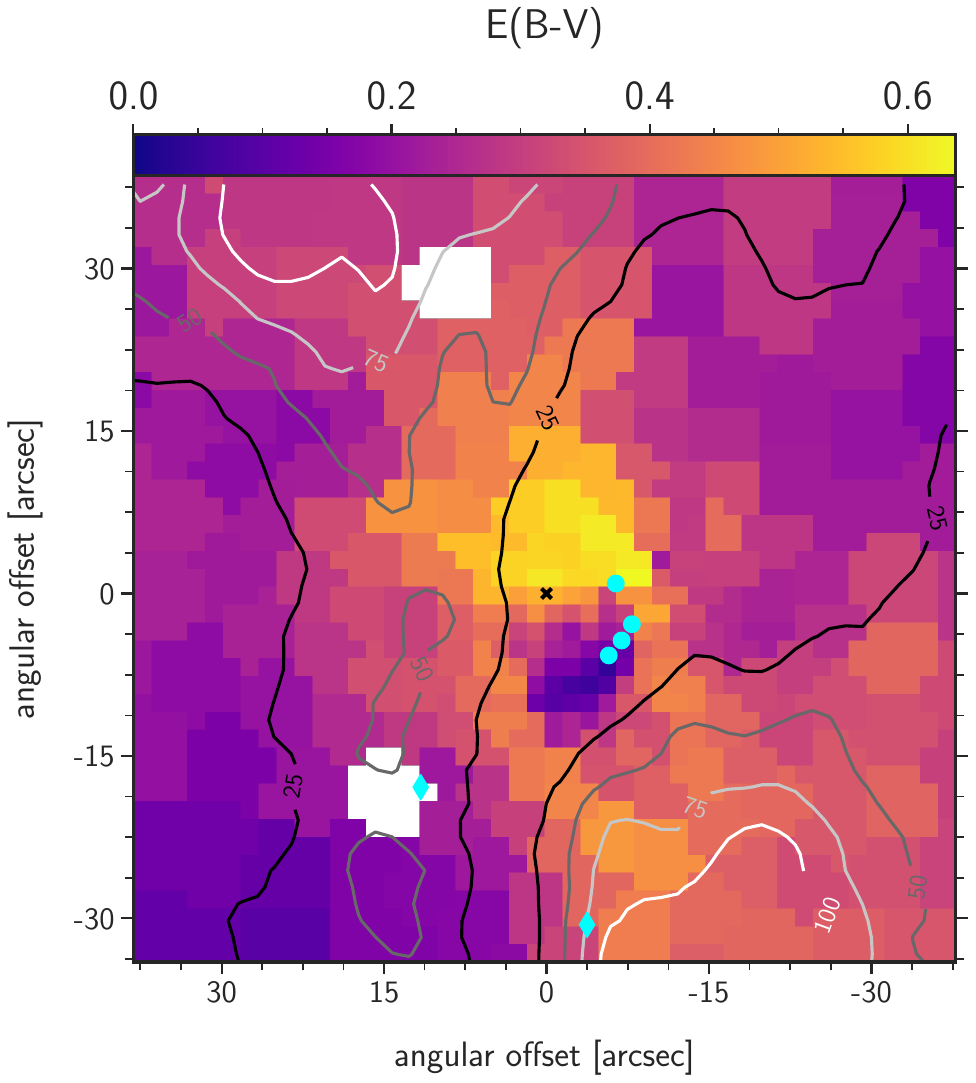}\medskip
	\caption{Enlarged reddening map of the central region of M83. The positions of six COS (HST) pointings of the YMCs studied by \citet{Hernandez2021} are indicated as cyan symbols. Circles correspond to clusters with very low \hi\ column densities and diamonds refer to somewhat higher column densities (see text). ALMA CO(2-1) emission line isocontours are overplotted in the top figure in five regular steps of 200 $\textrm{K} \cdot \textrm{km } \textrm{s}^{-1}$ starting at 100 $\textrm{K} \cdot \textrm{km } \textrm{s}^{-1}$. The bottom figures shows \hi\ 21 cm isocontours in four equal steps from 25 to 100 $ \textrm{Jy} \cdot \textrm{beam}^{-1} \cdot \textrm{ m}\textrm{ s}^{-1}$.}  \label{fig:redden2}
\end{figure}
% - - - - - - - - - - - - - - - - - - - - - - - - - - - - - - - - - - - - - - - - - 
\begin{figure}[htb!]
       \bigskip
       \bigskip
       \medskip
	\center \includegraphics[width=0.93\columnwidth]{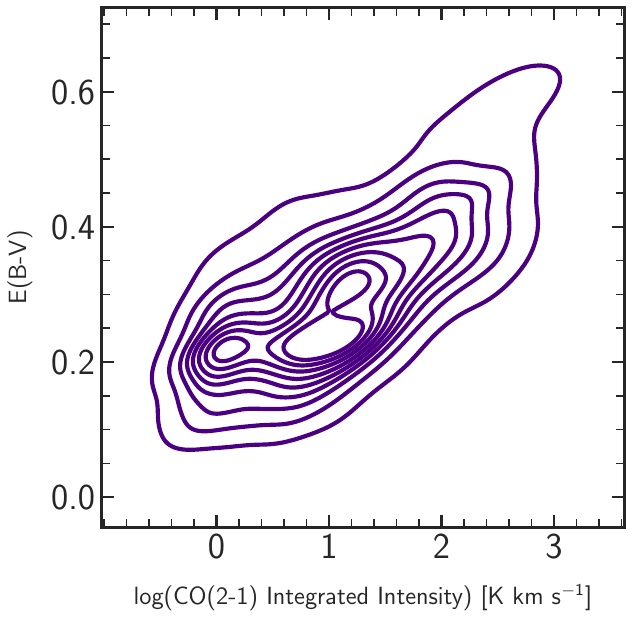}\medskip
	\caption{Isocontours showing the correlation between E(B-V) and CO(2-1) intensity in the central region of M83 shown in Fig.~\ref{fig:redden2}. }  \label{fig:corr_ebv_co}
\end{figure}

\subsection{Population synthesis method} \label{subsec:method}

Our population synthesis method has been described in detail in \citet{Sextl2023} and in Paper I. We repeat the basic aspects and also introduce a few changes relative to our previous work. The model spectrum M$_{\lambda}$ of the integrated stellar population of a galaxy surface element is calculated as a linear combination of spectra of single stellar populations (SSPs) of different ages and metallicities f$_{\lambda, i}$(t$_{i}$, [Z]$_{i}$)  with age t$_{i}$ and logarithmic metallicity [Z]$_{i}$ = log Z$_{i}$/Z$_{\sun}$ (Z$_{i}$ is the metal mass fraction)

\begin{equation}
  M_{\lambda} = D_{\lambda}(R_{V}, E(B-V)) \left[ \sum_{i=1}^{n_{SSP}} b_{i} f_{\lambda, i} (t_{i}, [Z]_{i}) + b_{a}f_{\lambda}^{a} \right].
\end{equation}  

The linear combination coefficients b$_i$ describe the contribution to the total spectrum of a burst with age t$_i$ and metallicity [Z]$_i$. $n_{SSP}$ is the number of SSPs used. We note that the observed and SSP template spectra are normalized to an average value of unity in the range of 5500 to 5550 \AA. Consequently, the sum over all b$_i$ is also equal to unity. The function D$_{\lambda}(R_{V}, E(B-V))$ accounts for the absorption by interstellar dust. We use the dust attenuation law by \citet{Calzetti2000} with variable total to selective extinction R$_V$. An introductory paragraph about the quantity R$_V$ is given in Sect.~\ref{subsec:redd}. We also account for the Milky Way foreground reddening adopting E(B-V)$_{MW}$ = 0.059 \citep{Schlafly2011} and the \citet{Cardelli1989} foreground reddening law with R$_V$\,=\,3.1. 

Our SSP fitting-templates are calculated from the Flexible Stellar Population Synthesis package (FSPS; version 3.2) \citep{Conroy2009, Conroy2010} using the MILES empirical library of stellar spectra \citep{Sanchez2006} augmented with a comprehensive set of spectra for young massive stars (see Paper I for details). They are identical to the SSPs in \citet{Sextl2023}. In total, we use a grid with 52 age and 9 metallicity values (468 SSPs in total). The age grid is in logarithmic steps, roughly one-third of the models are 20 Myr and younger, with the two youngest models corresponding to ages of 0.1 and 1.0 Myr (for a detailed description of the ages, see \citealt{Sextl2023}). The lowest metallicity value of the grid is -1.5 dex.

We note that M83 is extensively star-forming; it is therefore important to use a set of spectral templates that include very young populations (in the realm of several Myr) and stellar libraries with hot, massive stars. 

The coefficient b$_a$ accounts for a featureless AGN continuum with wavelength dependence $\lambda^{-0.5}$ \citep{Cardoso2017}, but its contribution is negligible. An AGN characteristic broad-line region is not identifiable in any of the centrally located spectra. 

As explained in \citet{Sextl2023} the b$_i$ are luminosity weighted fit coefficients. They can be expressed as

\begin{equation}
  b_{i} = N_{i} L_{i}(V).
  \label{eq:biNi}
\end{equation} 

L$_i(V)$ is the luminosity in the $V$ band of SSP isochrone \textit{i} before spectral normalization and N$_i$ is the number of stars represented by this SSP. We note that the N$_i$ are the actual number of stars multiplied by a common factor to account for the normalization. 

The metal content of the stellar population is obtained as follows. We use the current stellar mass of each SSP isochrone M$_i$, the metal mass fraction Z$_i$ = M$_{Z,i}$/M$_i$ and the mass-to-light ratio $\gamma_{i}$ = M$_i$/L$_i(V)$ and then calculate M$_Z$, the mass of metals contained by the stellar population 

 \begin{equation}
  M_{Z} = \sum_{i} N_{i}M_{i}Z_{i} = \sum_{i} b_{i}\gamma_{i}Z_{i}
  \end{equation}

and the total stellar mass

\begin{equation}
  M = \sum_{} N_{i}M_{i} = \sum_{i} b_{i}\gamma_{i}.
\end{equation}

For the metal mass fraction of the whole stellar population, we obtain

\begin{equation} \label{equation_Z}
  Z = {\frac{M_{Z}}{M}} = \sum_{i} \Tilde{b_{i}}Z_{i}
\end{equation}

with

\begin{equation}
  \Tilde{b_{i}} = {\frac{b_{i}\gamma_{i}}{\sum_{i} b_{i}\gamma_{i}}}.
\end{equation}

Finally, using Eq. (5) and (6) the logarithmic stellar metallicity is then given by

\begin{equation} \label{eq:Zlog}
  [Z] = log (Z/Z_{\sun}).
\end{equation}

We note that by using Eq. \ref{equation_Z} to \ref{eq:Zlog} we deviate from our earlier work in \citet{Sextl2023} and Paper I, where we have used luminosity-weighted means of [Z]$_i$ to calculate [Z]. We prefer the new approach because it provides a direct measurement of the metal content of the stellar population considered (the consequences of this new approach for the metallicities obtained in Paper I are discussed in Section \ref{sec:lightweightmetal}).

The logarithmic age of the total population is calculated by

\begin{equation}
    \log(t) = \sum_{i} \Tilde{b_{i}} \log(t_{i}).
\end{equation}

As mentioned above, M83 is a heavily star-forming galaxy. Thus, in order to investigate present star formation and chemical evolution and also to compare with independent spectroscopic observations of young stellar probes such as BSGs, SSCs and YMCs, as well as \hii\ regions, it is important to extract the properties of the very young stellar population from the population synthesis fit. This is done by introducing an SSP age limit t$_{lim}^{y}$ and separating the contribution of young stars through the condition t$_i$ $\le$ t$_{lim}^{y}$. 

The contribution of the young stars to the spectral fit is then

\begin{equation}
  b_{y} = \sum_{i_{y}} b_{i}, t_{i} \le t_{lim}^{y}.
  \label{by_def}
\end{equation} 

The average metallicity [Z]$_y$ and the age log(t$_y$) of the young population are calculated in a similar way as in the equations above, but with sums restricted to the age bin defined by t$_{lim}^{y}$.

% - - - - - - - - - - - - - - - - - - - - - - - - - - - - - - - - - - - - - - - - - 
\begin{figure}[htb!]
       \medskip
	\center \includegraphics[width=1\columnwidth]{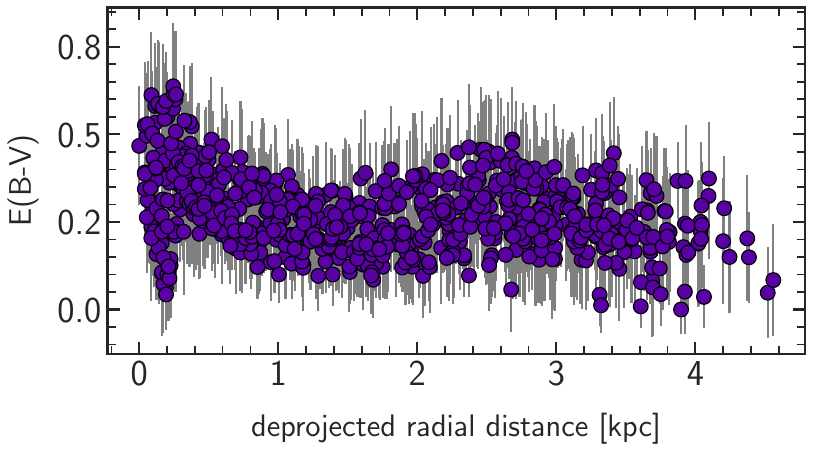}\medskip
    \center \includegraphics[width=1\columnwidth]{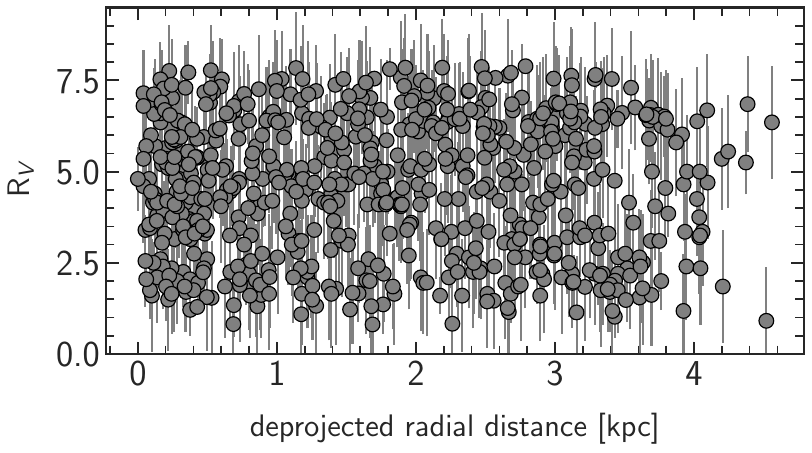}\medskip
    \center \includegraphics[width=1\columnwidth]{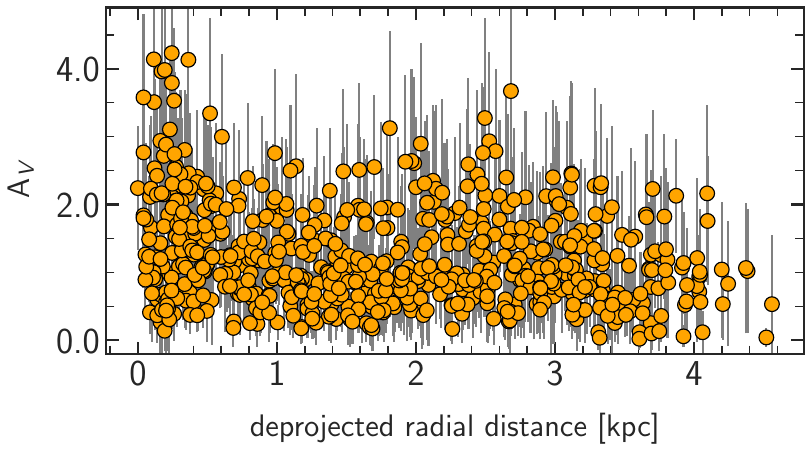}\medskip
	\caption{Radial distribution of E(B-V) (top), R$_V$ (middle) and A$_V$ (bottom).}     \label{fig:redgrad}
\end{figure}
%----------------------------------------------------------------------------------
\begin{figure}[htb!]
       \medskip
	\center \includegraphics[width=0.95\columnwidth]{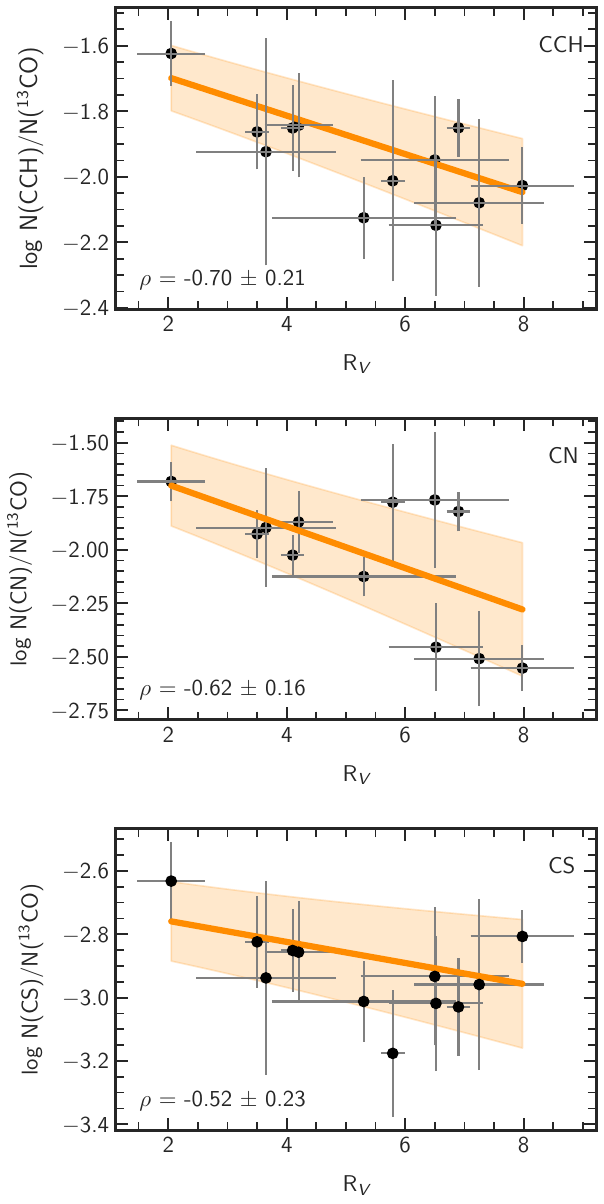}\medskip
	\caption{Correlation of R$_V$ with column density ratios of three molecular species,  CCH (ethynyl radical, top), CN (cyano radical, middle), and CS(carbon monosulfide, bottom), at selected positions in the central region of M83. A linear regression fit is shown in orange colors. The Pearson correlation coefficient $\rho$ is printed at the left bottom of each plot and uncertainties of each data point are shown.}     \label{fig:rvcorr}
\end{figure}

% - - - - - - - - - - - - - - - - - - - - - - - - - - - - - - - - - - - - - - - - - 
\begin{figure}[htb!]
       \smallskip
	\center \includegraphics[width=0.95\columnwidth]{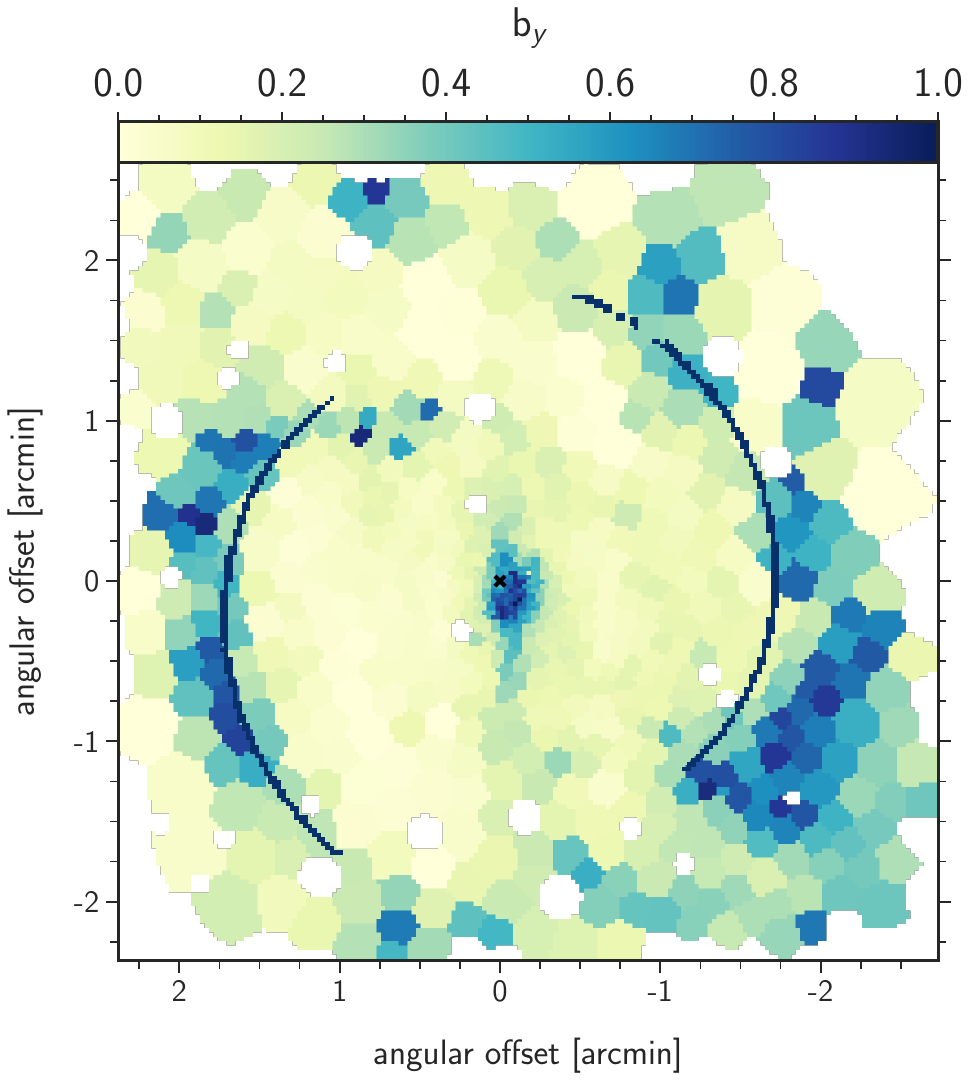} 
	\caption{Spatial maps of the fit contribution coefficients b$_y$ for the young stellar population. The plot indicates which fraction of the measured flux at 5500~\AA\ comes from stars younger than 0.1 Gyr. The location of the CO spiral arm fit is shown in dark blue.}   \label{fig:youngold}
\end{figure}

The star formation rate of the young population is

\begin{equation}
    \psi_{y} = {M_{iso}\cdot L(V) \over t_{lim}^{y}} \sum_{i_y} {b_i \over L_i},
\end{equation}

where M$_{iso}$ is the original stellar mass of each SSP isochrone, when all its stars were formed. M$_{iso}$ is the same for each SSP. L(V) is the observed V band luminosity of the TYPHOON Voronoi bin.

We will use t$_{lim}^y$ = 0.1 Gyr to determine the ages and metallicities of the young stellar population. For the calculation of star formation rates we will adopt 20 and 5 Myr, respectively (see Section \ref{subsec:sfr}). To distinguish between these two cases, we introduce the star formation rates $\psi_y^{20}$ and $\psi_y^{5}$ accordingly.

In a similar way as for the young stars, we can also define the old stellar population. We introduce an SSP age limit t$_{lim}^o$ and consider the contributions of old stars through the condition t$_i \ge$ t$_{lim}^{o}$. This allows us to calculate the fit contributions b$_o$, metallicities [Z]$_o$, and ages log(t$_o$) as in Eq. (9) to (12), but with sums over SSP that satisfy the age criterion for the old population. As a typical value for t$_{lim}^{o}$, we will use 1.6 Gyr (see Paper I and \citealt{Sextl2023}).

The analysis requires a certain minimum signal-to-noise ratio for the observed spectra. Depending on the galaxy surface brightness we may need to combine spectra of adjacent individual spaxels to obtain this level. We accomplish this by Voronoi binning using \citet{cappellari2003} until we obtain a minimum signal-to-noise ratio of 60 in the stellar continuum at 5000 \AA. The binning is performed over the FOV shown in Figure \ref{fig:Vband}, and this spatial scale is maintained in the subsequent 2D maps, except for the zoomed-in view of the center.
Because of the small distance of M83 and the high surface brightness, most of our Voronoi bins are relatively small and allow us to trace spatial variations such as spiral arms, except in the very outer galaxy regions where we start averaging over too large spatial dimensions. In those outer regions we exclude bins consisting of more than 330 TYPHOON spaxels. This leaves us with spectra of 737 bins distributed throughout the galaxy. Figure \ref{fig:redden1} gives an impression of the spatial distribution of the bins. 
With a distance of 4.81 Mpc, M83 is rather close and bright point sources could possibly interfere with the fitting procedure of unresolved stellar populations. We remove those areas following the foreground star removal algorithm outlined in \citet{Clark2018}, utilizing the Python package PTS-7/8 \footnote{https://github.com/SKIRT/PTS} (a python toolkit for the SKIRT radiative transfer code presented in \citet{Camps2015, Camps2020}). The procedure operates in our case on a CTIO 1.5m $R$-band image of M83 and starts by retrieving relevant entries from the 2MASS All-Sky Catalog of Point Sources \citep{Cutri2003} using the Astroquery Python library\footnote{https://github.com/astropy/astroquery}. For each cataloged position, the algorithm examines a small image patch, subtracting the estimated background and searching for a local peak using the Photutils Python package\footnote{https://github.com/astropy/photutils}. The process ignores positions where no reasonable signal-to-noise (SN$>$5) peak is found within a few pixels' radius. To prevent removal of compact sources within target galaxies (such as \hii\ regions and massive clusters), the procedure only considers peaks that show no deviations from being true point sources, except for potential saturation in the brightest stars. We find that for TYPHOON IFU data with an FWHM of 2 arcsec for the point spread function \citep{Agostino2018} and a low spatial resolution of 1.65 arcsec, this approach is good enough. 

We apply a radial velocity correction to the observed spectra and then focus on the analysis of their stellar contribution by masking out spectral regions contaminated by ISM emission or absorption lines. However, we account for the nebular continuum emission potentially underlying the stellar spectra. The strength of this continuum is determined from the strength of observed H${\alpha}$ or H${\beta}$ emission. For details, we refer to \citet{Sextl2023}. 

After these preparative steps, the coefficients b$_i$ and b$_a$ are determined by adopting a grid of R$_V$ and E(B-V) value. For each pair of these quantities we redden the model spectra with D$_{\lambda}(R_{V}, E(B-V))$ and apply the bounded variable least-squares algorithm (BVLS, \citealt{Stark1995}, see also \citealt{Sextl2023}) to directly solve for the coefficients b. We then use the resulting model spectrum and calculate a $\chi^2$ value by comparing with the observed spectrum. The minimum of $\chi^2$ defines the best fit with respect to R$_V$ and E(B-V). We estimate errors by fitting again the observed spectra but this time modified by adding Monte Carlo Gaussian noise with zero mean and a standard deviation corresponding to the flux error at each wavelength point. The uncertainties of the stellar population parameters are then calculated as the standard deviation of their distributions produced by 50 such Monte Carlo realizations.

% - - - - - - - - - - - - - - - - - - - - - - - - - - - - - - - - - - - - - - - - - 
\begin{figure}[htb!]
    \medskip
	\center \includegraphics[width=0.95\columnwidth]{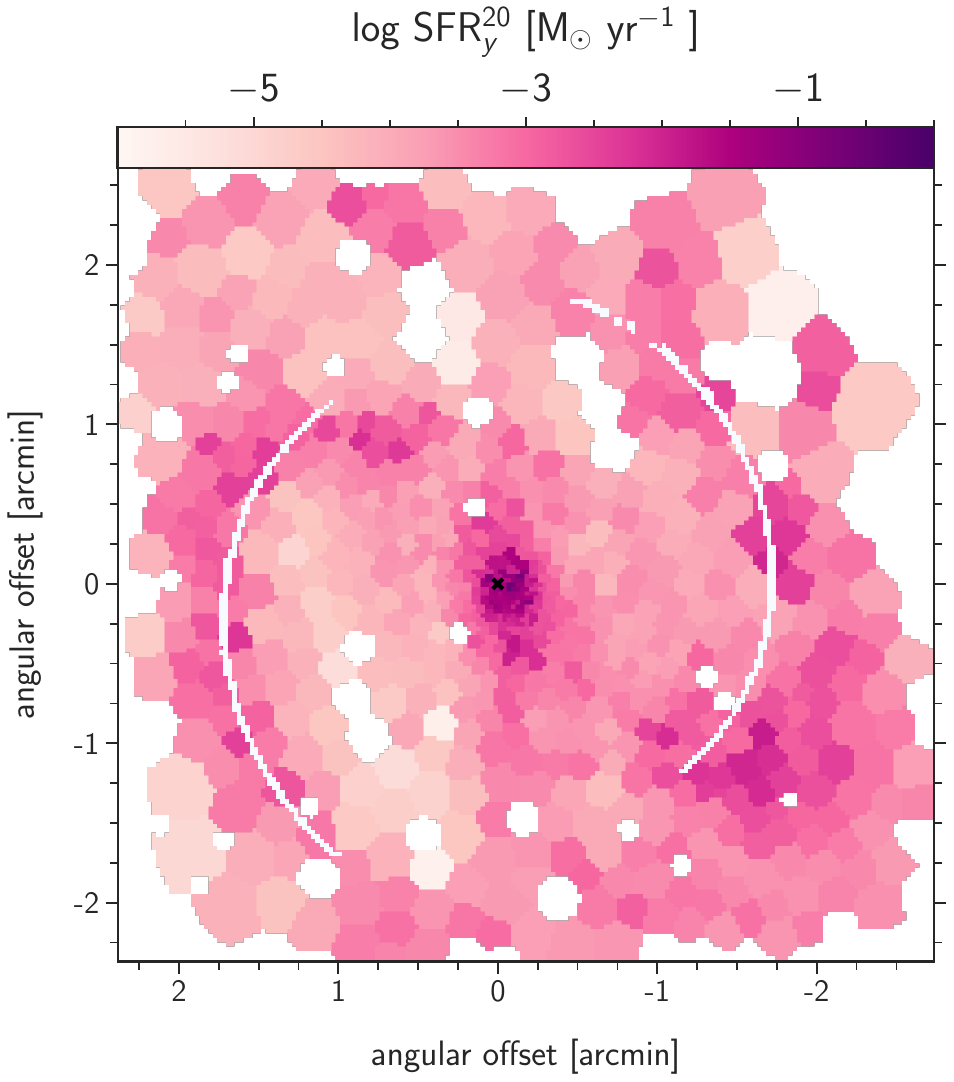}\bigskip
	\caption{Spatial map of the star formation rate of the young stellar population $\psi_y^{20}$ over the last 20 Myr.} \label{fig:sfr}
\end{figure}

\begin{figure}[htb!]
\medskip
	\center \includegraphics[width=0.95\columnwidth]{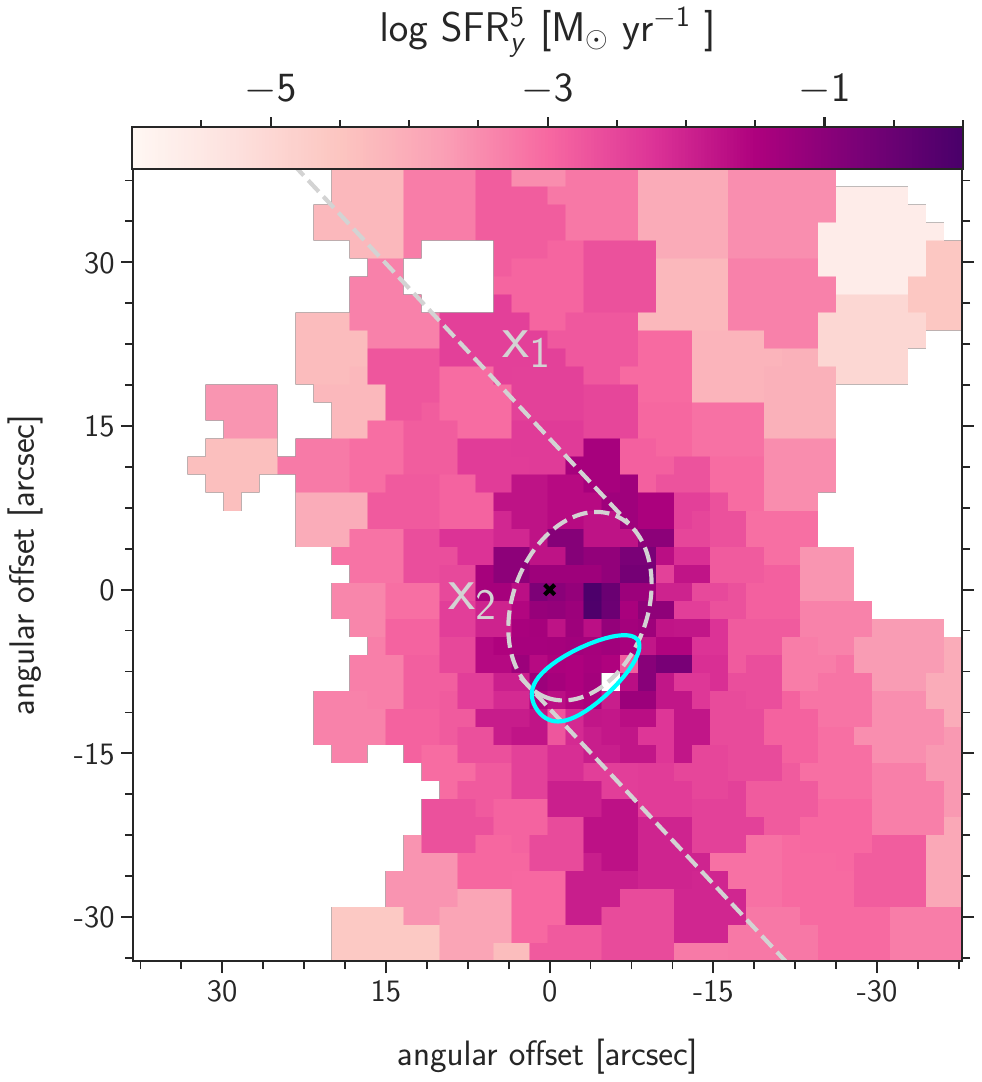}\medskip
    \center \includegraphics[width=0.95\columnwidth]{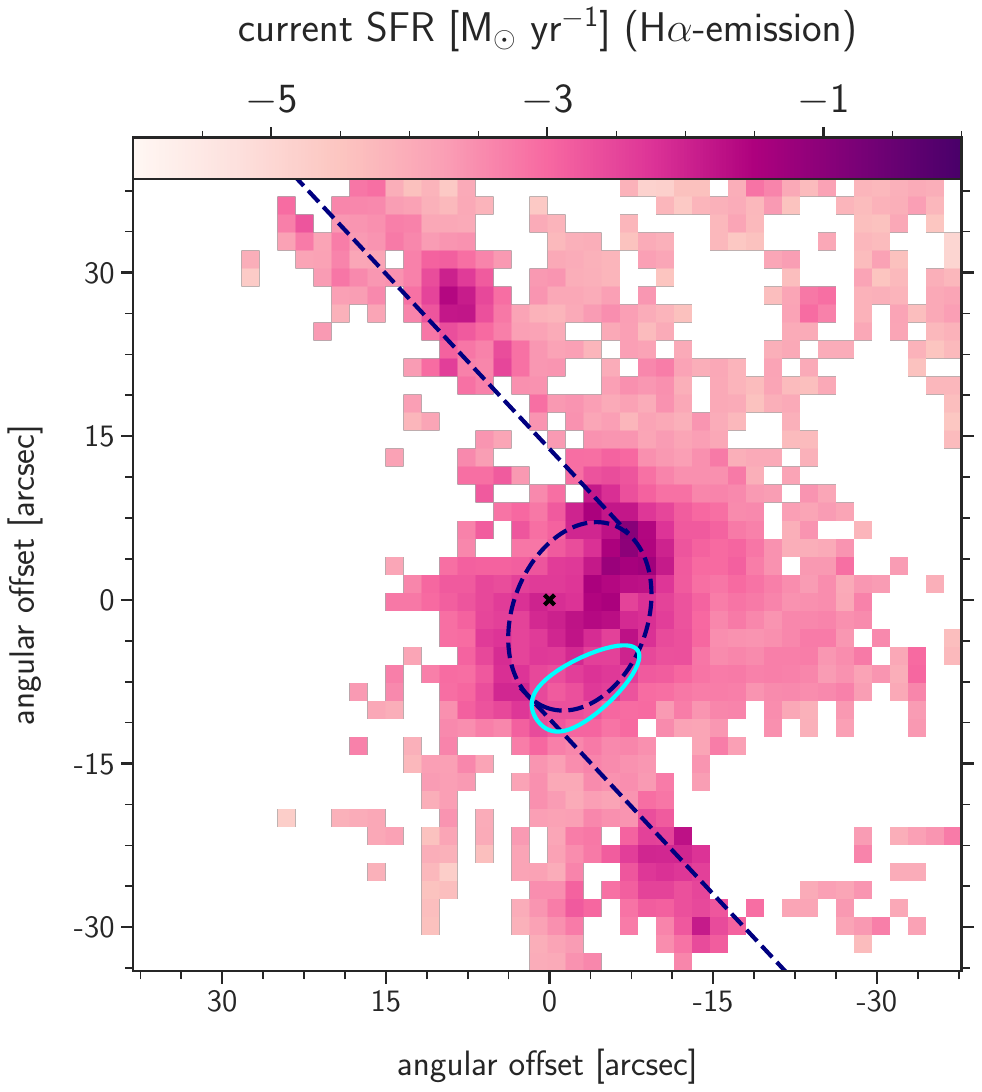}\medskip
	\caption{Top: Central map of most recent star formation $\psi_{y}^{5}$ encountered over the last 5 Myr as obtained from our population synthesis analysis. Bottom: Central map of the star formation rate derived from H$\alpha$ emission. The position of the dust cavity (in aqua) and the gas flow along potential x$_1$ \& x$_2$ orbits are also shown (see text).}     \label{fig:sfr_halpha}
\end{figure}
% - - - - - - - - - - - - - - - - - - - - - - - - - - - - - - - - - - - - - - - -

In the following sections, we present the results of our M83 population synthesis analysis.
% - - - - - - - - - - - - - - - - - - - - - - - - - - - - - - - - - - - - - - - - - 
\begin{figure}[htb!]
       \medskip
	\center \includegraphics[width=0.95\columnwidth]{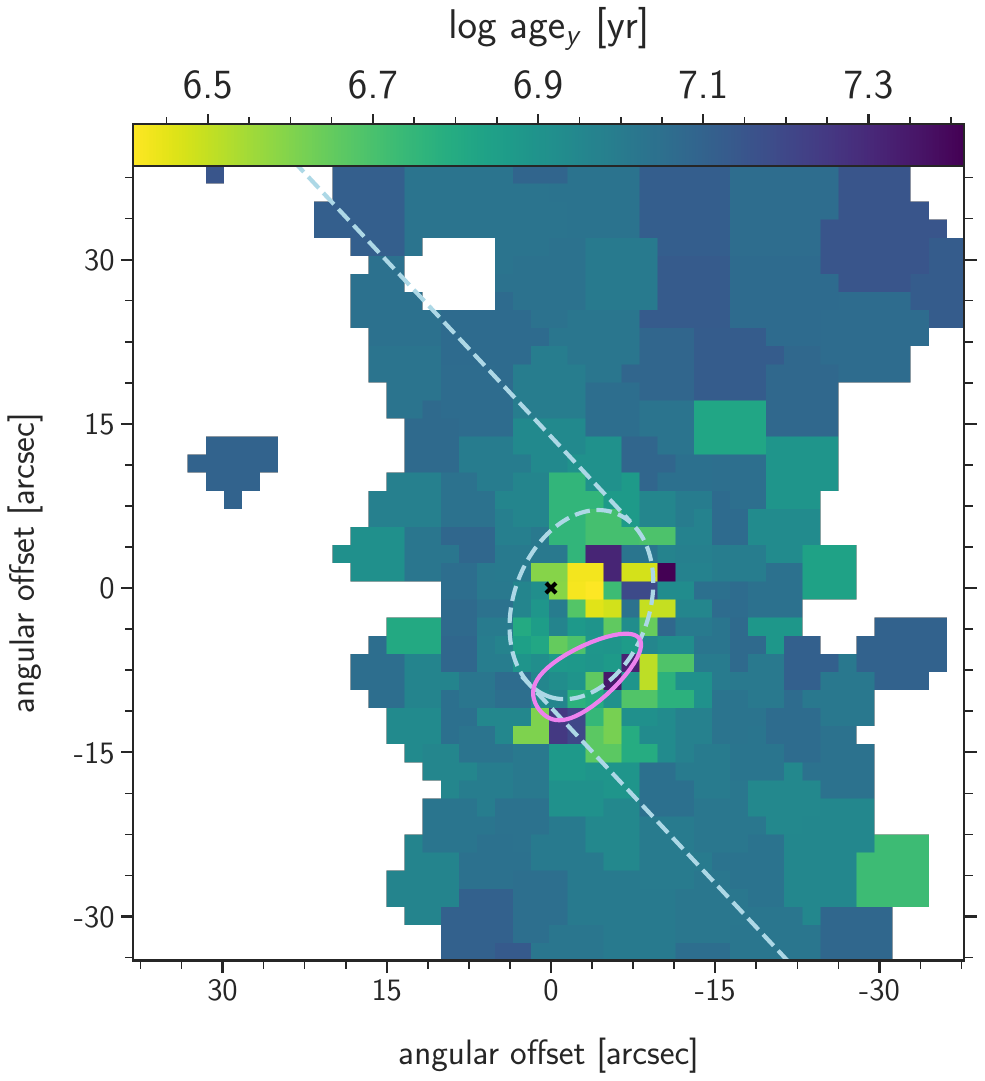}\medskip
	\caption{Central map of the average age (in years) of the young stellar population. x$_1$ \& x$_2$ orbits (light blue) and the dust cavity (pink) are again overlaid.}     \label{fig:ageyoung}
\end{figure}
% - - - - - - - - - - - - - - - - - - - - - - - - - - - - - - - - - - - - - - - - - 
\begin{figure}[htb!]
       \medskip
	\center \includegraphics[width=0.95\columnwidth]{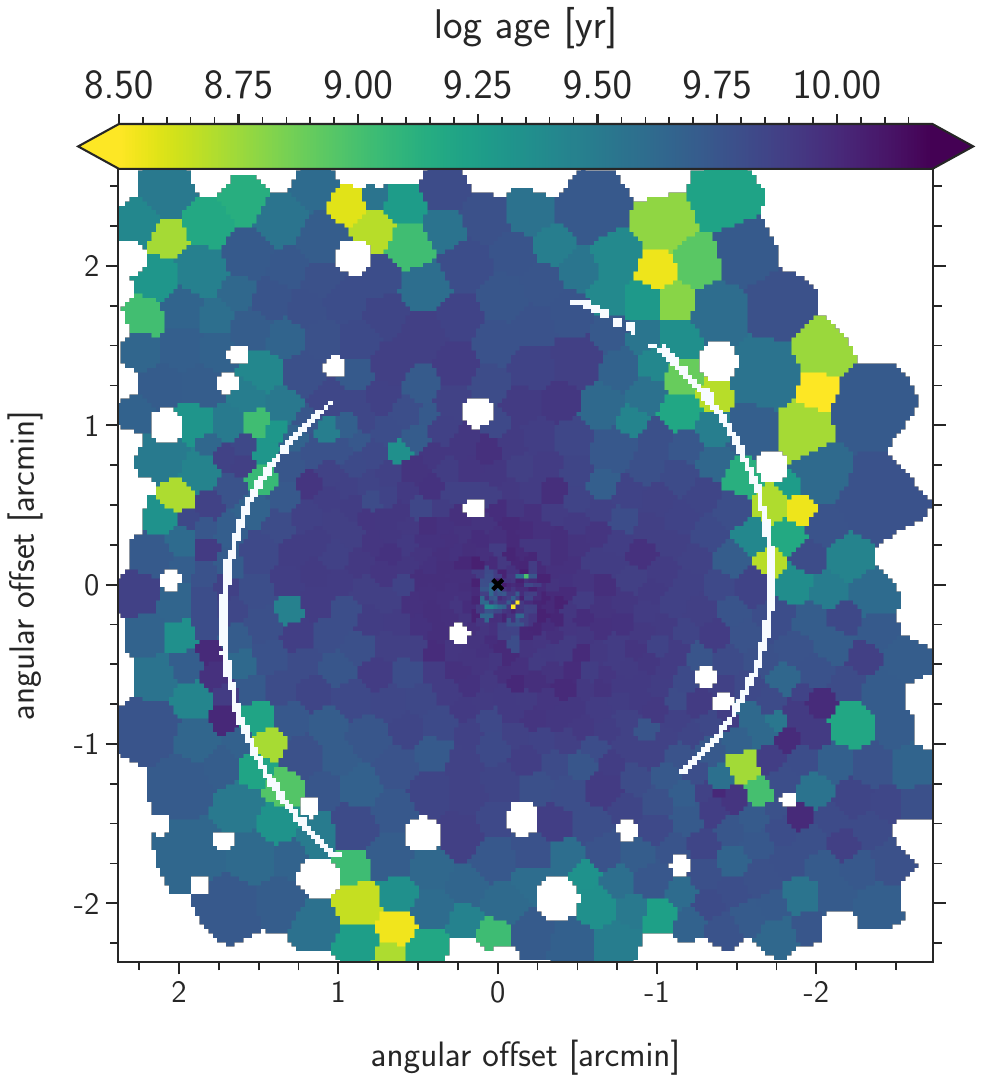}\medskip
    \center \includegraphics[width=0.95\columnwidth]{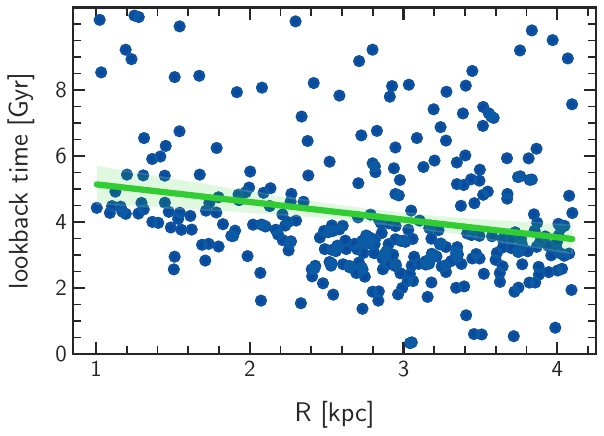}\medskip
	\caption{Top: Map of the average stellar population age. CO spiral arm fits are over-plotted for orientation. Bottom: Lookback time when 85 percent of the mass of the total stellar population has formed as a function of galactocentric radius in kiloparsec. A linear regression curve (in green) is also shown. A trend decreasing with radius indicates inside-out growth.}    \label{fig:agegrad}
\end{figure}

% - - - - - - - - - - - - - - - - - - - - - - - - - - - - - - - - - - - - - - - - - 
\begin{figure*}[htb!]
       \medskip
	\center \includegraphics[width=0.8\textwidth]{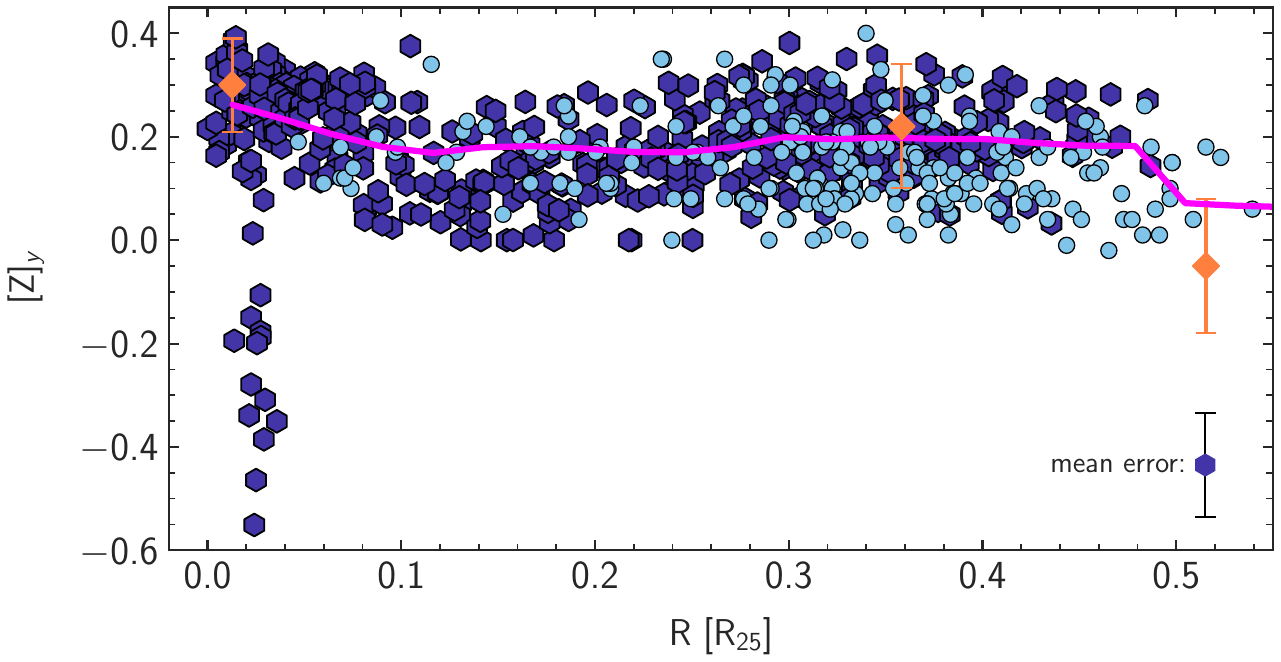}\medskip
	\caption{The radial gradients of the young population metallicities (in dark blue hexagons) from full-spectral fitting compared with results from individual \hii\ regions. The \hii\ region strong-line abundances (in light blue) were determined from the same TYPHOON data cube (\citealt{Grasha2022}, see text). In addition, \hii\ region abundances obtained with the direct method by \citet{Bresolin2005, Bresolin2016} are shown in orange. Errorbars of individual Voronoi bins and TYPHOON \hii\ regions are omitted for better visibility. They have average values of 0.1 dex as indicated. The prediction of the chemical evolution model presented in \citet{Bresolin2016} is shown in pink. For discussion see text.}     \label{fig:zgradyoung}
\end{figure*}

\section{Results: Reddening, star formation, ages, and inside-out growth} \label{sec:results}

In the subsections following below we present the results with respect to reddening, extinction, star formation, average ages, and inside-out growth of the galactic disk. 

\subsection{Reddening and extinction: a dust cavity in the center and correlations with molecular gas} \label{subsec:redd}

Figure \ref{fig:redden1} shows the reddening map of M83 obtained with our population synthesis approach. The contribution of Milky Way foreground reddening is subtracted as described above. Generally, we find enhanced reddening close to the spiral arms and the bar, which coincides with an increased emission from atomic and ionized H, cold molecular CO gas and cold and warm dust (see, for instance, \citealt{Frick2016}). For the northern spiral arm, where we have encountered a difference between the H${\alpha}$ and CO logarithmic fit, the maximum of the dust absorption follows the molecular gas.

%In addition, we note an interesting discrepancy between the fit of the northern spiral arm of Figure \ref{fig:Halpha} and the amount of reddening. Here, we find an obvious dent in the spatial dust distribution. This can also be very clearly seen in the integrated intensity map of CO(1-0) in \citet{Koda2023} and in the maps in \citet{Frick2016}. 

Figure \ref{fig:redden2} provides an enlarged reddening map of the central region of the galaxy and reveals a complex structure of the dust distribution. Reddening is enhanced in a confined area north of the center, whereas southwest of the center we find a region of low extinction corresponding to a dust cavity. Its estimated size, with a diameter of 11 arcseconds, corresponds to approximately 260 parsecs. This dimension is comparable to the Local Bubble observed in the Milky Way \citep{Zucker2022}, hinting at a similar origin related to past enhanced star formation with supernovae. As we shall see later, the cavity is almost void of molecular and atomic ISM gas. \citet{DellaBruna2022} also highlight a significant dip in velocity dispersion within this area ($\sigma < 60$ km/s).

The dust cavity is also strikingly apparent in the multicolor optical images obtained with the Hubble Space Telescope. Figure 1 in \citet{Harris2001} very clearly reveals a bright elongated region south of the center that exactly coincides with our cavity. Very obviously, a hole in the dust distribution allows us to peek into a region containing bright massive clusters and stars. Measurements of the Balmer decrement through H${\alpha}$ and H${\beta}$ imaging (Figure 4 in \citealt{Harris2001}) indicate very low reddening in our cavity region. We take this as confirmation that full-spectral fitting is a very powerful tool for investigating the distribution of reddening and interstellar dust.

We note that the complex pattern in the center of M83 with an extinction cavity surrounded by large extinction is not unique. Figure 3 of the TYPHOON study by \citet{Sextl2023} reveals a similar structure for the barred spiral NGC~1365. Because of the significantly larger distance of this galaxy (18 Mpc) the spatial resolution is worse and the effect appears to be less striking, but it is definitely there. 

As the comparison with the ALMA CO (J = 2-1) isocontours shown in the upper part of Figure \ref{fig:redden2} indicates, the distribution of molecular gas roughly coincides with the distribution of dust. The peak of CO emission correlates with the northern region of strong reddening, and the dust cavity shows a low level of CO emission. Figure \ref{fig:corr_ebv_co} documents the overall correlation between dust and CO in the center. This supports a scenario where the molecular gas forms out of the atomic gas in regions of enhanced dust density with dust grain surfaces acting as a catalyst. The ALMA observations used in the figures are described in \citet{Hirota2018} and were downloaded from the PHANGS archive\footnote{\url{https://sites.google.com/view/phangs/home/data}}. 

Comparing with the  \hi\ 21 cm isocontours of the THINGS survey \citep{Walter2008}, we indeed find that the regions with the maximum concentration of E(B-V) and CO coincide with very low neutral hydrogen gas concentration (Figure \ref{fig:redden2}, bottom part). This seems to agree with the transformation of atomic to molecular gas in the presence of dust. On the other hand, our dust cavity with low CO emission is still located in an area of very low \hi\ column densities. This is further confirmed by the work of \citet{Hernandez2021}, who carefully studied \hi\ column densities in the central region by analyzing the strength of Ly${\alpha}$ absorption lines superimposed on the UV spectra of the integrated light of YMCs. The location of these YMCs is also shown in Figure \ref{fig:redden2}. The four objects of low absorption column densities (log[N(HI)/cm$^2$] $\le$ 20) are located close to or inside the dust cavity. This points to an overall decrease of ISM matter density of all components, neutral and cold molecular gas and dust, in this part of the center. This is corroborated by \citet{Hernandez2023} in their JWST mid-IR investigation of the center of M83 detecting molecular emission of cold and warm H$_2$ gas. 

The radial distribution of E(B-V), R$_V$, and A$_V$ is shown in Figure \ref{fig:redgrad}. The radial plot of E(B-V) reflects the complex structure in the center and a reddening decline towards an outer average level of E(B-V) $\approx$ 0.2 mag. The small increase around 2.5 kpc is caused by the contribution of spiral arms and the large areas of related star formation. As in our study of NGC 1365, we find a wide range of R$_V$ values, while A$_V$ roughly follows the reddening distribution. No covariances between E(B-V) and R$_V$ were encountered.

In general, the R$_V$ parameter plays a significant role in characterizing the shape of the extinction or attenuation curves and can potentially provide information on the nature of dust grains. For a typical extinction situation with individual bright stellar point sources such as supergiant stars shining through a dust screen, sight lines that include larger dust particles create an extinction curve with a high value of R$_V$ and produce flatter or 'greyer' extinction curves \citep{Cardelli1989, Fitzpatrick2007}. This means that the extinction is more uniform across different wavelengths, resulting in less wavelength-dependent reddening. A large variety of R$_V$ values is observed in the literature, for instance, R$_V \sim 5$ in selected molecular clouds, R$_V \sim 1 - 3$ for supernovae type Ia and R$_V \sim 2 - 7$ for massive blue supergiant stars \citep{Mandel2011,Dwek2005,Urbaneja2017, Kudritzki2024}. 

Unfortunately, for our population synthesis analysis with stellar sources distributed within the dusty ISM the interpretation is more complicated because we are dealing with attenuation curves \citep{Calzetti2013}. Still, with our comprehensive information on the spatial distribution of R$_V$ it is particularly interesting to correlate its values with other properties in M83. In the radial plot (Figure \ref{fig:redgrad}, middle), the values appear to be arbitrarily distributed. We also find no correlation with metallicities, stellar ages, or the obtained color excess values in the associated bins. However, in the special case of M83, we also have the unique opportunity to compare with the results of an astrochemical ISM study. \\
\citet{Harada2019} analyzed ALMA data in the center of M83 and obtained the column densities of different molecular ISM species. This was done at 12 selected pointings in and around the circumnuclear ring (see their Figure 1). As the authors give the coordinates of these positions, we were able to obtain the corresponding R$_V$ values from our fit. The main results are shown in Figure \ref{fig:rvcorr}. We find an indication of an anti-correlation between our R$_V$ and their fractional abundances of CCH (ethynyl radical) and somewhat weaker dependencies with CN (cyano radical) and CS (carbon monosulfide). A regression fit with uncertainty bands and the corresponding Pearson correlation coefficients with error estimates resulting from Monte Carlo bootstrapping are also plotted.

In other words, the molecular species CCH (and, to a lesser extent, CN and CS) might be preferentially abundant in regions where smaller dust grains dominate the dust size distribution. Of course, this interpretation of the observed correlations assumes that R$_V$ and dust grain size remain related in the population synthesis attenuation analysis. We also note that the abundances of additional ISM species listed in Tables 4-6 in \citet{Harada2019} (including CH3OH, H2CO and NNH) show no correlation with the slopes of the extinction curves we obtain.\\
To further explain these findings, it is crucial to understand the multifaceted role of dust in the ISM. First, the surfaces of dust grains act as catalysts for chemical reactions, particularly in the formation of molecular hydrogen and complex organic molecules \citep{Wakelham2017, Herbst2009}. Second, dust particles absorb stellar radiation, especially in the ultraviolet and at optical wavelengths, and re-emit this energy in the infrared and sub-millimeter range in the form of thermal emission \citep{Viaene2016}. Third, dust grains participate in heating of the ISM via the photoelectric effect (where they absorb UV photons and eject energetic electrons) as well as through collisions with gas particles \citep{Draine1978}.\\ 
The efficiency of these dust-mediated processes is significantly influenced by the size distribution of dust grains. As noted previously, an increase in R$_V$ shifts the grain size distribution towards larger grains. This shift has several important consequences. Larger grain sizes result in reduced dust opacity, particularly in the UV range, allowing deeper penetration of far-ultraviolet (FUV) radiation into molecular clouds \citep{Weingartner2001, Abel2008}. This leads to elevated FUV radiation energy densities and accelerated gas heating \citep{Wolfire2022}. 
%\citet{Soliman2024} demonstrated through extensive magnetohydrodynamic simulations that this can potentially lead to the rapid shutdown of star formation due to the lack of available molecular hydrogen. \\

On the other hand, FUV photons from neighboring massive stars lead to so-called photo-dissociation regions (PDRs) at the edges of molecular clouds.
%At the edges of molecular clouds, the so-called photo-dissociation regions (PDR) located, where far-ultraviolet (FUV) photons from neighboring massive stars influence the gas properties in this regime.
It is known that in PDRs the ISM species CCH and CN are predominantly produced through photochemical processes \citep{Martin2015}. CS formation is also favored in PDR environments via ion reactions \citep{Martin2008,Forrey2018}. \citet{Roellig2013} demonstrated using CLOUDY calculations that larger grains lead to hotter PDRs and consequently lower column densities of H$_2$, CO, and OH$^-$ compared to scenarios with smaller grains. Although the authors did not explicitly test this for other ISM components, it seems plausible that this trend extends to gas-phase-created species such as CCH, CN, and CS \citep{Sipila2020}. This would explain the observed correlations. In that sense, Figure \ref{fig:rvcorr} provides a strong motivation to further investigate the complex interaction between dust grain size and the chemical composition of the ISM.  
% - - - - - - - - - - - - - - - - - - - - - - - - - - - - - - - - - - - - - 
\begin{figure}[htb!]
       \medskip
    \center \includegraphics[width=1\columnwidth]{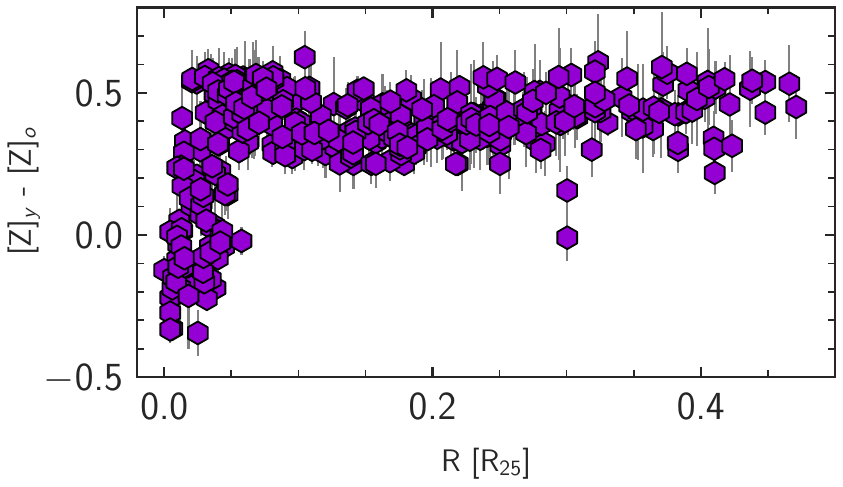}\medskip
	\caption{The radial gradient of the metallicity difference between the young and old stellar populations. This time, the error bars are explicitly shown.}     \label{fig:DeltaZ}

\end{figure}

\subsection{The Spectral Contribution of the Young and Old Population}

In our population synthesis fit we explicitly separate between the contribution of a young and older population to the integrated stellar spectrum. The weights b$_y$, as defined in Eq.~\ref{by_def}, describe the relative contribution of the young population to the spectral fit. Their distribution across the galaxy is shown in Figure \ref{fig:youngold} (we have adopted t$_{lim}^{y}$ = 0.1 Gyr for the plot). We find a strong contribution close to the spiral arms and in the center. The inter-arm region, on the other hand, is clearly dominated by older stars. We note that the contribution of the young stars is slightly shifted away from the spiral arms marked by CO and E(B-V) (see Figure \ref{fig:redden1}). This may be the result of disk rotation, spiral pattern speed, and star formation and will certainly be worth further investigation.

The uncertainties of our measurements of the metallicity and age of the young population depend on b$_y$. The dependence is weak as long as the weights are greater than 0.1. However, once b$_y$ becomes smaller than 0.1, the uncertainties increase and we will not include the corresponding values of metallicities and ages in our discussion and plots. Similarly, the uncertainties for the old population depend on b$_o$ and we will not include the results obtained from the spatial bins with b$_{o} \le 0.2$. Errors will be indicated in the plots in the following sections.

\subsection{Star formation} \label{subsec:sfr}

Figure \ref{fig:sfr} provides spatial maps of $\psi_{y}^{20}$, the star formation rate of the young stellar population. We have chosen a maximum age of $t_{lim}^{y}$ = 20 Myr in order to capture the most recent star formation events (see, for instance, \citealt{Riffel2021}). We clearly see enhanced star formation close to the spiral arms and a giant active star formation region in the southwest of the galaxy at the leading egde outside the spiral arm. We also encounter the well-known nuclear star burst of M83 with strongly enhanced star formation. 

In Figure \ref{fig:sfr_halpha} we display enlarged maps of the star formation rates in the center. We show the results obtained from full spectral fitting and from H$\alpha$ flux measurements. For the former, we have chosen $t_{lim}^{y}$ = 5 Myr for reasons explained below. We also include orbits of matter infalling along the bar and around the circumnuclear ring reconstructed from the CO maps shown in \citet{Harada2019}. The x$_1$ orbit types are elongated and aligned with the major axis of the bar and drawn as straight lines in our figures. x$_2$ orbits, in contrast, are perpendicular to the bar's major axis, often rather circular and typically found inside the Inner Lindblad Resonance (ILR) of the galaxy \citep{Contopoulos1989}. They both play important roles in the galactic structure and in the evolution of the bar. \citet{Harada2019} argue that at the intersection of both orbit types, molecular gas clouds accumulate and collide, eventually triggering star formation events. 

The orbital period of a typical x$_2$ orbit is 10 Myr \citep{Harada2019}. We therefore select t$_{lim}^y$ = 5 Myr for the determination of star formation rates. This is also a time scale that agrees with the ages of the very young stars in the center (see \ref{subsec:ages}). Extensive star formation is visible throughout the central region within the x$_2$ orbits. However, we do not see a clearly pronounced minimum of star formation in the region of the dust cavity and the void of molecular gas.

For the sake of testing the method of full-spectrum fitting, it is interesting to compare our star formation results based on stellar population synthesis against the standard method using H$\alpha$ emission. We use the \citet{Kennicutt1998} relationship between the star formation rate and (dereddened) H$\alpha$ fluxes in the version updated by \citet{Calzetti2007}.  We plot the central distribution of star formation rates derived from the H$\alpha$ fluxes in Figure \ref{fig:sfr_halpha} (bottom). We exclude spaxels with signal-to-noise ratios less than 3 in H$\alpha$, H$\beta$, [NII]6583 or [OIII]5007 and regions which lie above the Kewley line \citep{kewley2001} in the BPT diagram (i.e.~being photoionized by shocks) and we remove the effects of reddening and attenuation of the H$\alpha$ fluxes by applying the Balmer decrement method and by additional correction for Milky Way foreground extinction.\\
In general, the distribution of the SFR derived from H${\alpha}$ in the ring and the x$_2$ orbits is very similar to the structure obtained with population synthesis for the last 5 Myr. We find again very strong star formation activity inside the x$_2$ orbits. Enhanced star formation activity is also found along the x$_1$ orbits.  

% - - - - - - - - - - - - - - - - - - - - - - - - - - - - - - - - - - - - - - - - - 
\begin{figure}[htb!]
       \medskip
	\center \includegraphics[width=0.95\columnwidth]{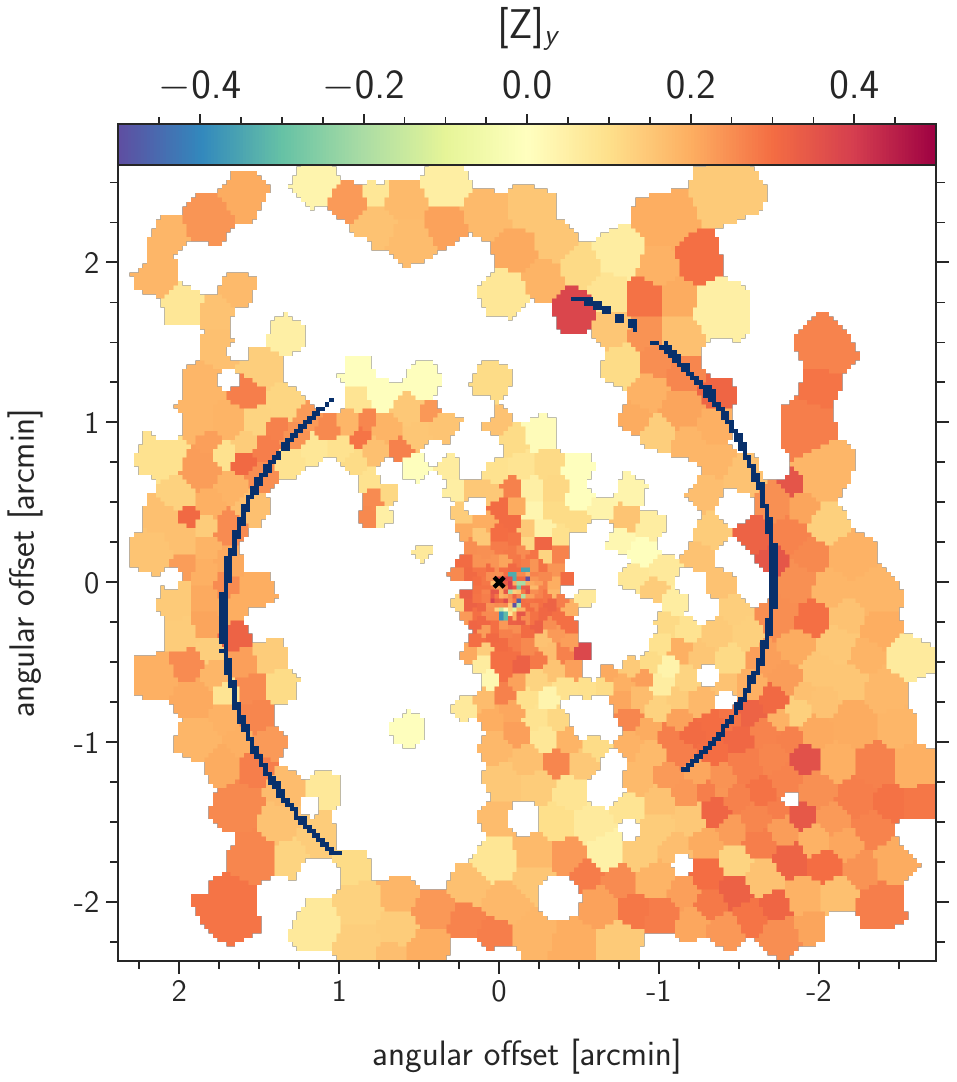}\medskip 
	\caption{Metallicities of the young stellar population. CO spiral arms are indicated as before. Regions with $b_y<0.1$ were removed in order to exclude results with larger uncertainties (see text).}    \label{fig:agezmap}
\end{figure}

% - - - - - - - - - - - - - - - - - - - - - - - - - - - - - - - - - - - - - - - - - 
\begin{figure}[htb!]
       \medskip
	\center \includegraphics[width=0.95\columnwidth]{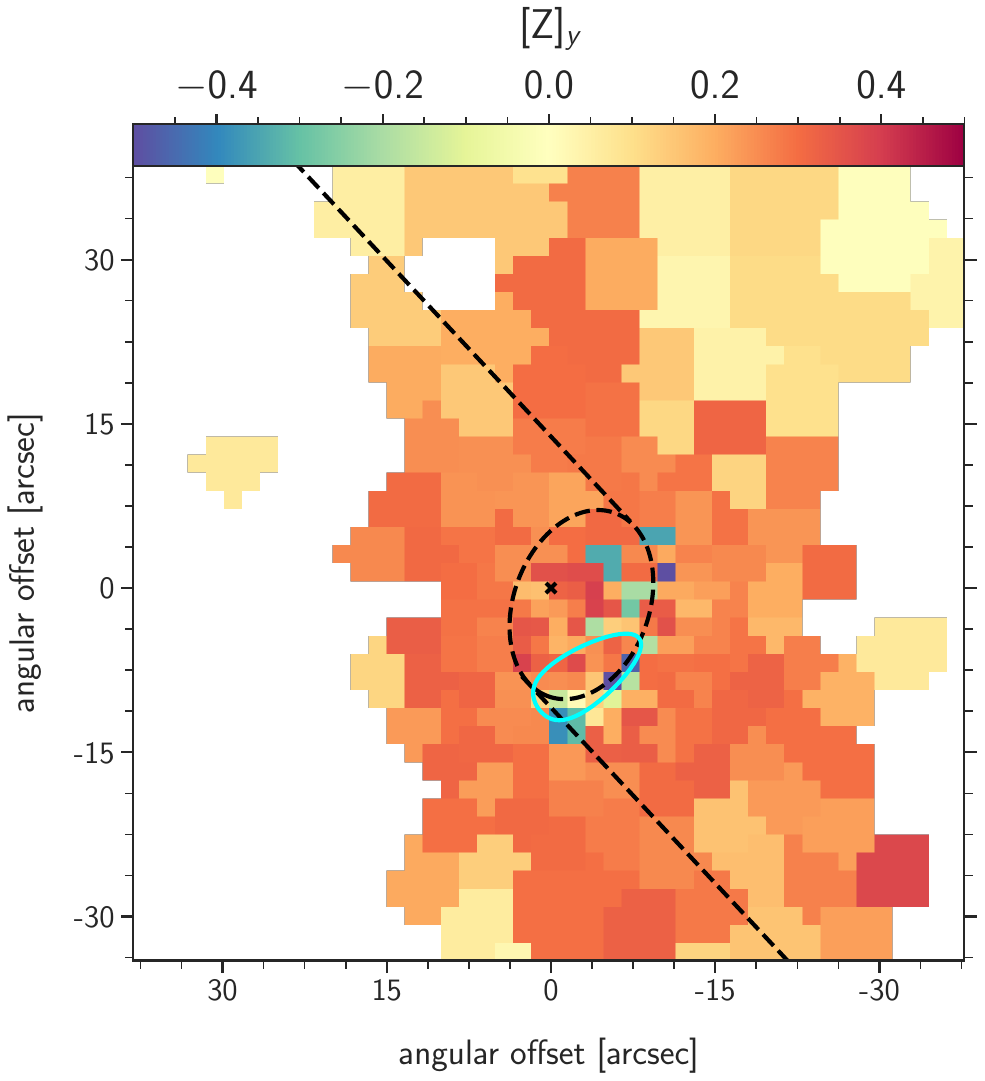}\medskip
	\caption{Enlarged map of [Z]$_y$ around the center of the galaxy. }    \label{fig:zmaplarge}
\end{figure}
% - - - - - - - - - - - - - - - - - - - - - - - - - - - - - - - - - - - - - - - - - 
%\begin{figure}[htb!]
%       \medskip
%    \center \includegraphics[width=0.95\columnwidth]{figure17b_ZoDIFF.pdf}\medskip
%	\caption{Central map of the metallicity difference [Z]$_y$ -[Z]$_o$ between the young and old stars.}    \label{fig:zoldcenter}
%\end{figure}
% - - - - - - - - - - - - - - - - - - - - - - - - - - - - - - - - - - - - - - - - - 

\subsection{Stellar ages: the young population in the center} \label{subsec:ages}

\begin{figure}[htb!]
       \medskip
	\center \includegraphics[width=1\columnwidth]{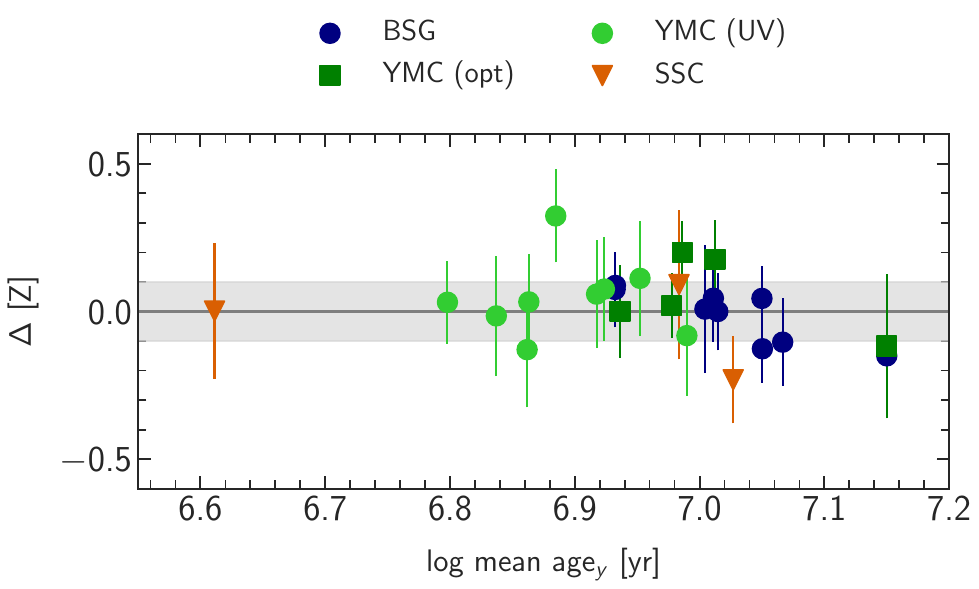}\medskip
    \center \includegraphics[width=1\columnwidth]{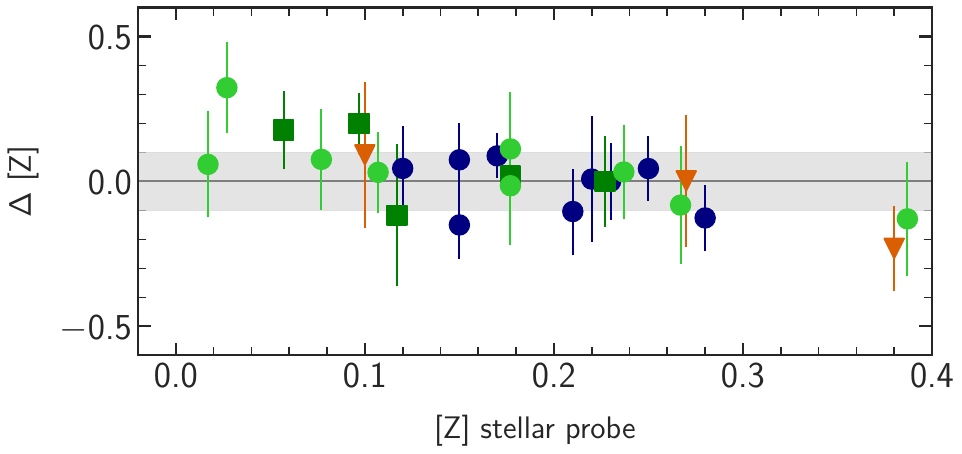}\medskip
    \center \includegraphics[width=1\columnwidth]{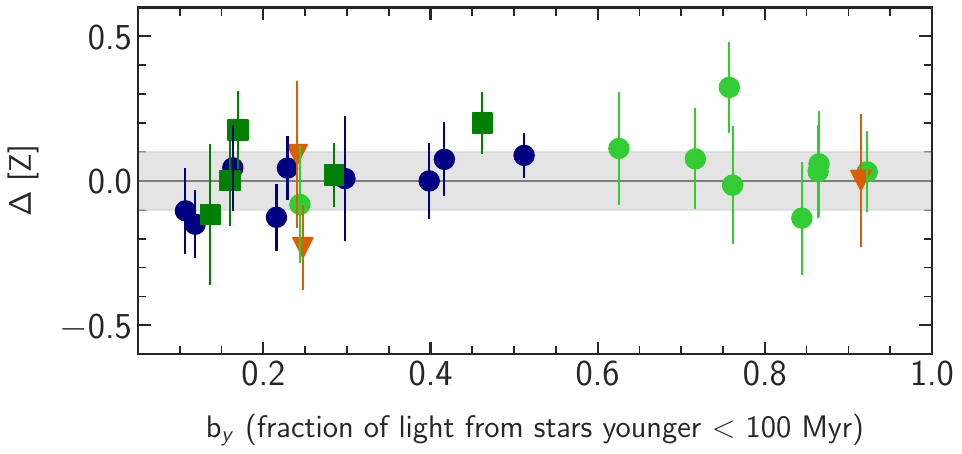}\medskip
	\caption{Difference $\Delta$[Z] = [Z]$_{y}$ - [Z]$_{probe}$ between population synthesis and individual stellar source metallicities as a function of the mean flux-weighted age (top), [Z$_y$] (middle) and b$_y$ (bottom). Color coding: dark blue circles: BSG, red triangles: SSC, dark green squares: YMC (optical analysis), light green circles: YMC (UV analysis). Errors result from the addition of stellar source and population synthesis errors in quadrature. The shaded gray strip indicates a difference $\le$ $\pm$0.1 and is added for orientation.}   \label{fig:stellar_probes}
\end{figure}

Figure \ref{fig:ageyoung} displays the average age of the young stellar population in the center of M83. We have chosen t$_{lim}^{y}$ = 0.1 Gyr for the calculation, but the mean ages are much younger. We note that the region of maximum star formation and the largest dust content is dominated by very young stars with ages smaller than 10 Myr. The regions with extremely young stars of 5 Myr age coincide with very strong H${\alpha}$ emission. The dust cavity region and the eastern part of the x$_2$ orbit harbor slightly older stars with ages around 7 Myr. Typical age uncertainties are 1 Myr. Thus, we think the age difference is significant. We note that the ages obtained with our population synthesis approach are in good agreement with the spectroscopic study by \citet{Bresolin2002} of nuclear hot spots around the center of M83.

More generally, in a wider region around the center, the population is still very young with an age smaller than 20 Myr. This is in strong contrast with the outer galactic disk, as we shall see in the next subsection. 

\subsection{Inside-out growth of the galactic disk}

As discussed in Paper I the average age of the stellar population gives a good overall impression about the evolution of the galactic disk. Figure \ref{fig:agegrad} (top) shows the age map of the total stellar population of M83. We find a strong contribution by young stars in the star-forming regions close to the spiral arms. The remaining parts of the disk show a strong contribution from older stars. Their ages do not seem to vary much. This is consistent with \citet{Pessa2023}, who found older stellar populations evenly distributed throughout galactic disks of PHANGS-MUSE galaxies. In M83, there seems to be a slight increase when leaving the center, with a maximum at approximately 0.15 R/R$_{25}$ and then the ages drop again very moderately. As we have found in Paper~I for the barred spiral NGC~1365, this might be indicative of galactic inside-out growth. We, therefore, use the same method as in Paper I (Eq.~9) to calculate the lookback time when 85 percent of the mass of the total stellar population has formed. The result is shown in the bottom part of Figure \ref{fig:agegrad}, which indicates a trend, with outer stars being on average slightly younger. The regression shows a gradient in the time of disk formation of $-0.53 \pm 0.13 $ Gyr/kpc or $-4.8 \pm 1.2$ Gyr/R$_{25}$. This is comparable, given the limits of accuracy, to $-4.4 \pm 1.3$ Gyr/R$_{25}$ found in Paper~I for~NGC 1365. The inside-out growth may have proceeded similar in both galaxies. 

\section{Results: Stellar Metallicity} \label{sec:results_2}

\subsection{Radial metallicity gradients} \label{subsec:gradients}

In a first step, we discuss the metallicity of the young population defined by ages smaller than t$_{lim}^{y}$ = 0.1 Gyr. We display the radial distribution in Figure \ref{fig:zgradyoung} and find two striking features. In the innermost disk and the center, we encounter a significant drop of [Z]$_y$. As we shall see in the next subsection, these lower values of metallicity are confined to a small region in the center of the galaxy. We note that we have encountered a very similar behavior in NGC 1365 (see Paper I). 

The distribution further out in the disk (R\,$\ge$\,0.04~R$_{25}$) is stunningly flat and barely shows a negative metallicity gradient. We notice a weak drop at 0.15 R/R$_{25}$ and then an increase toward smaller and larger radii. A regression fit yields [Z]$_y$ = 0.20($\pm$0.02) $-$ 0.01($\pm$0.06) R/R$_{25}$. For comparison, we have also added the radial distribution of \hii\ region oxygen abundances in units of the solar value \citep{Asplund2009} to our plot of [Z]$_y$. The \hii\ region data were obtained by \citet{Grasha2022} from the M83 TYPHOON survey data cube with the use of HIIphot \citep{Thilker2002} for identification and LZIFU \citep{Ho2016} for the measurement of emission line strengths. We have utilized these line strengths and applied the \citet{Dopita2016} strong line calibration for the estimate of oxygen abundances. They confirm the very shallow gradient in this part of the disk, which was already noted by \citet{Grasha2022}. For completeness, we also add the \hii\ region metallicities obtained by \citet{Bresolin2005, Bresolin2016} with the direct method using weak auroral emission lines and with uncertainties smaller than 0.3 dex.

The flat metallicity distribution in the radial range 0\,$\le$\,R/R$_{25}$\,$\le$ 0.5 was predicted by the chemical evolution model presented in \citet{Bresolin2016}. The model is based on the method by \citet{Kudritzki2015} and adopts constant ratios of the galaxy mass loss and the accretion mass gain to the star formation rate. It uses the azimuthally averaged ratios of stellar mass to gas mass for the calculation of the radial metallicity profile. The prediction of this model is shown in Figure \ref{fig:zgradyoung} as well. The metallicity profile simply reflects the average radial ratio of stellar to gas mass. It reproduces the observed metallicity distribution very well, but it does not explain the spatial bins of low metallicity in the center. We note that our IFU study presented here covers only the inner 4 kpc of the galactic disc. The spectroscopic work presented in \citet{Bresolin2016} shows very clearly that the metal content drops towards larger galactocentric distances in a huge extended disk with a metallicity 0.4 dex below solar.

In Figure \ref{fig:DeltaZ} we show the metallicity difference \mbox{[Z]$_y$ $-$ [Z]$_o$} between the young and old stellar populations. The average value of 0.4 dex in the disk indicates a significant degree of chemical evolution. We note that the distribution is flat, which means that the old population also does not exhibit a gradient. We also note a strong increase in the center. We attribute this to the decline of [Z]$_y$ in the center. We will continue to discuss this further in the next subsection.

% - - - - - - - - - - - - - - - - - - - - - - - - - - - - - - - - - - - - - - - - - 

\begin{figure*}[htb!]
\centering
  \begin{minipage}[t]{0.48\textwidth}
    \includegraphics[width=\textwidth]{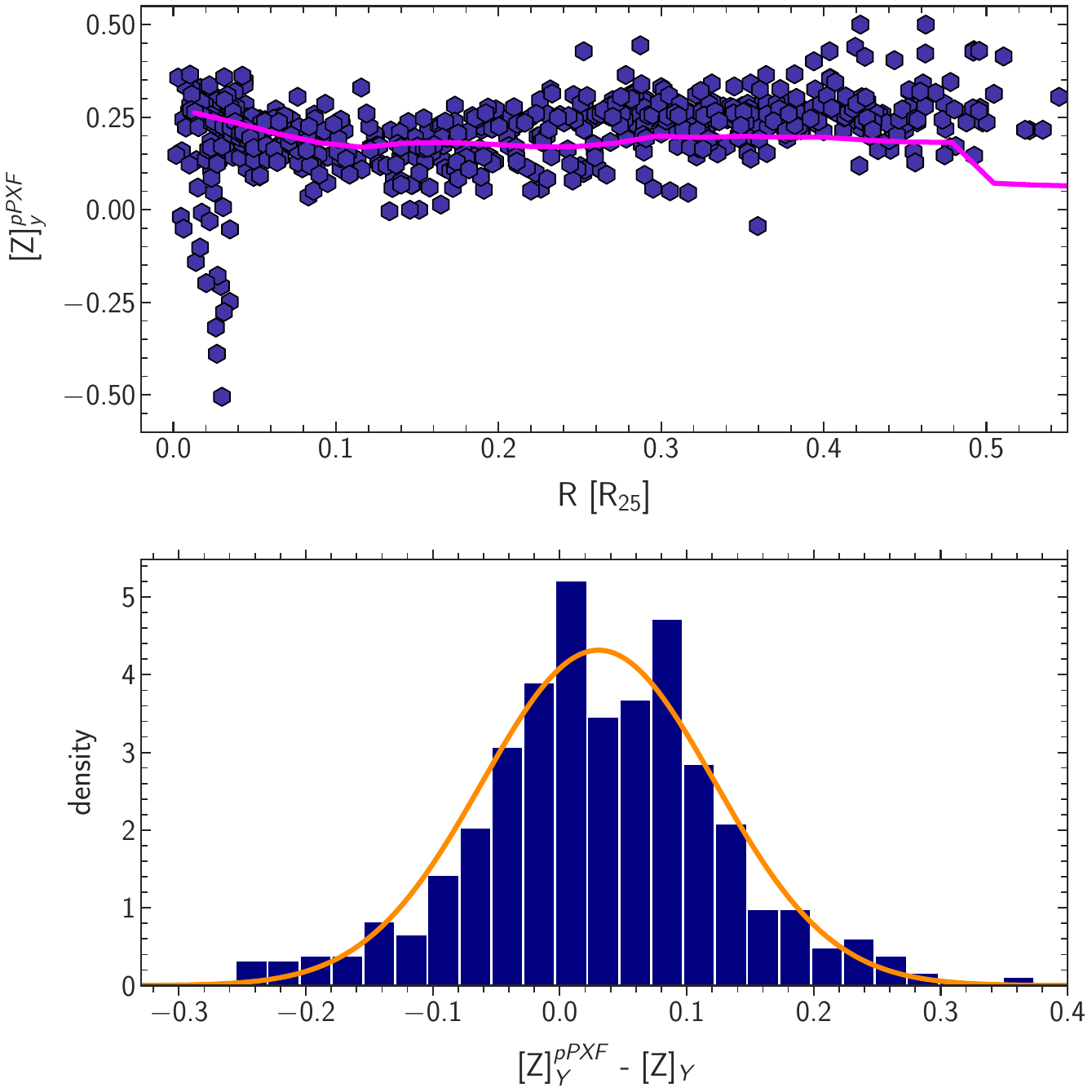}
  \end{minipage}
  \hfill
  \begin{minipage}[t]{0.46\textwidth}
    \includegraphics[width=\textwidth]{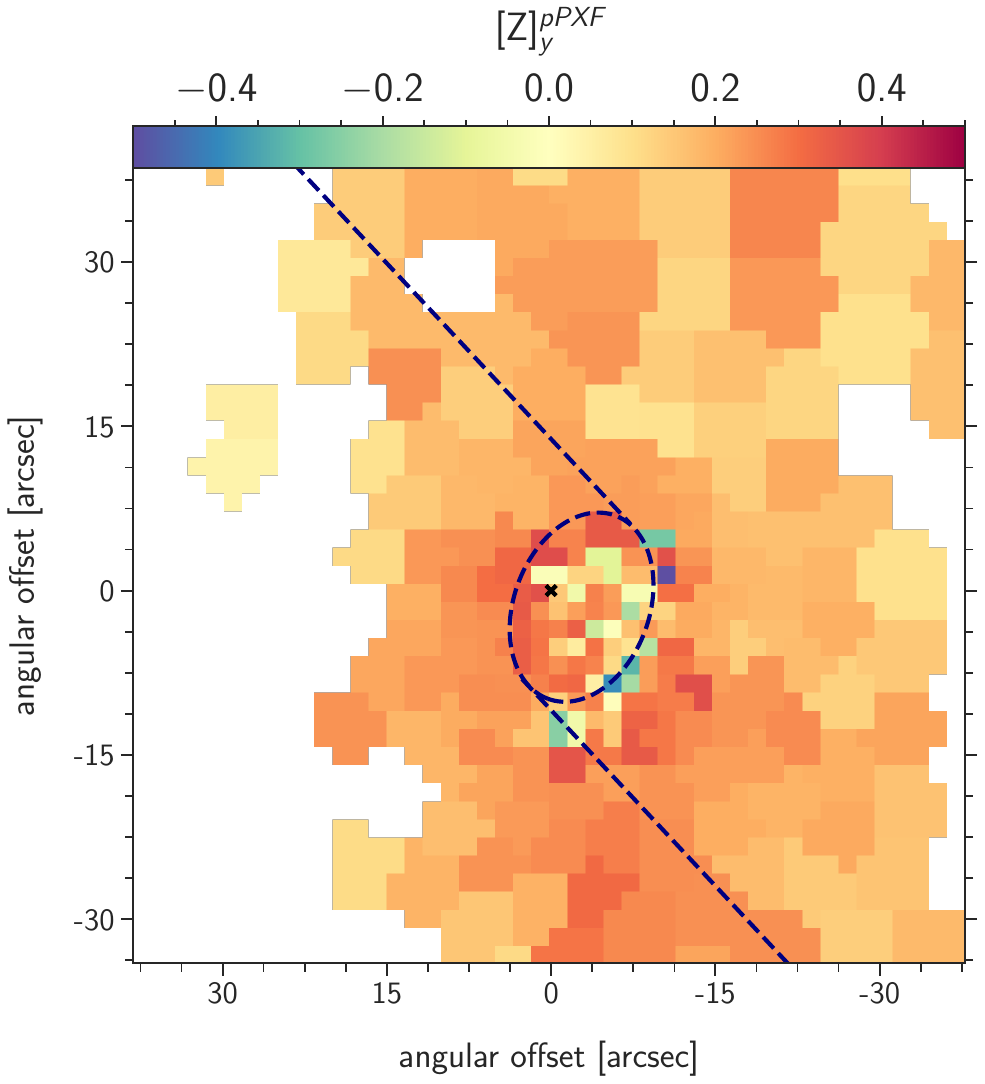}
  \end{minipage}
  \caption{Population synthesis with pPXF. Top left: Radial gradient of [Z]$_y^{pPXF}$, the chemical evolution model from \citet{Bresolin2016} is shown again in pink to guide the eye. Bottom left: Distribution function of the difference [Z]$_y^{BVLS}$ $-$ [Z]$_y^{pPXF}$, the orange line is a fitted Gaussian function. Right: Central map of [Z]$_y^{pPXF}$}.   \label{fig:pPXFfit}
\end{figure*}

\subsection{ Galactic distribution of metallicity} \label{subsec:spiral}

One of the major purposes of this work is the investigation of deviations from simple gradients and from azimuthal homogeneity of the chemical composition of the young stellar population. In Figure \ref{fig:agezmap} we show the corresponding spatial map of metallicity. Although the analysis of the result is hampered by the large areas with missing information due to b$_{y} \le 0.1$, we can still see a pattern. We see increased metallicity close to the spiral arms, the bar, and around the center together, with a drop between the center and the spiral arms. In addition, the star-forming area southwest of the western spiral shows increased metallicity. We do not encounter regions of lower metallicity in the downstream areas of the spiral arms, as detected in the case of the \hii\ region analysis of the two grand-design barred spirals NGC 1365 and NGC 2997 by \citet{Ho2017, Ho2018} and \citet{Kreckel2019}. Our result agrees with the \hii\ region study by \citet{Chen2024}. In the case of a galactic disk without a metallicity gradient, stellar population synthesis fitting is very likely not sensitive enough to detect these very small subtle variations.

In the following, we discuss the metallicity distribution in the center. Figure \ref{fig:zmaplarge} provides an enlarged map of the [Z]$_y$ distribution in the central region of the galaxy. We find significantly reduced metallicities in the western region of the x$_2$ orbits where the bar-related x$_1$ orbits hit the circumnuclear ring. Most striking is the southern part, where low-metallicity spatial bins align exactly along the ring just at the edge of the dust cavity, while the metal content in the dust cavity is higher. 

As in Paper I we have carried out detailed tests to check whether the lower [Z]$_y$ are an artifact of the fitting procedure or caused by numerical uncertainty. We compared the minimum spectral fit values $\chi^2$ as a function of [Z]$_y$ and found no systematic difference. Keeping R$_V$ fixed to the Calzetti standard value of 4.05 produces equally low metallicity. In addition, as we show in Section \ref{sec:approaches}, applying an independent alternative population synthesis algorithm yields a similar result. We also note that the comparison with the results of detailed spectroscopy of individual stellar sources carried out in Section \ref{subsec:compare} indicates that our metallicity diagnostic of [Z]$_y$ is reliable.

We attribute the lower metallicities encountered to the matter inflow of metal-poor gas from the circumgalactic medium leading to the presence of young metal-poor stars. An alternative is interrupted chemical evolution where star formation is stopped by activity of the central AGN for several Gigayears and then resumes with gas ejected by stellar winds from earlier generations of stars. This scenario has been discussed in detail in Paper I. We also note the most interesting spatial high-resolution 3D near-infrared spectroscopic study by \citet{Diaz2006}. This work concludes that a recent dwarf galaxy-like interloper has affected the central region of M83. We plan to investigate the very complex situation in the center of M83 (see also \citealt{Houghton2008,Knapen2010,DellaBruna2022}) in follow-up work. 

Such peculiar star-forming rings/disks with unexpected metallicity are a recurring observation in the literature. Our findings in NGC 1365 and NGC 5236 align with several other examples, contributing to a growing body of evidence for this phenomenon. \citet{Rosado2020} find hints for this in the MUSE data of NGC~1300 and NGC 1097. \citet{Robbins2025} describes the central regions of NGC 5806 as having a "counter-intuitive stellar population" which could be linked to the presence of the AGN and gas flows. NGC 5728 is another example found by \citet{Shimizu2019} who speculate about a recent minor merger as an explanation. Further examples include NGC 7552 \citep{Seidel2015} as well as several galaxies in the MUSE-Timer sample (including M83) \citep{bittner2020,Gadotti2019}. These observations share common characteristics: they occur in barred spiral galaxies and were detected with full-spectral fitting techniques on IFU data. However, we note that these studies do not separately investigate the young stellar population. They also apply luminosity-weighted (or mass-weighted) averages of logarithmic metallicity (see Section \ref{sec:lightweightmetal}). A follow-up study using our methodology will be very interesting.

%Figure \ref{fig:zoldcenter} shows the metallicity difference [Z]$_y$$-$[Z]$_o$ between young and old stars in the central region of M83. We find a well-confined area with values close to zero or even negative. This is caused by the central drop in [Z]$_y$, but also by an increase in [Z]$_o$ in the central bulge. 

% - - - - - - - - - - - - - - - - - - - - - - - - - - - - - - - - - - - - - - - - - 
\begin{figure}[htb!]
       \medskip
	\center \includegraphics[width=0.95\columnwidth]{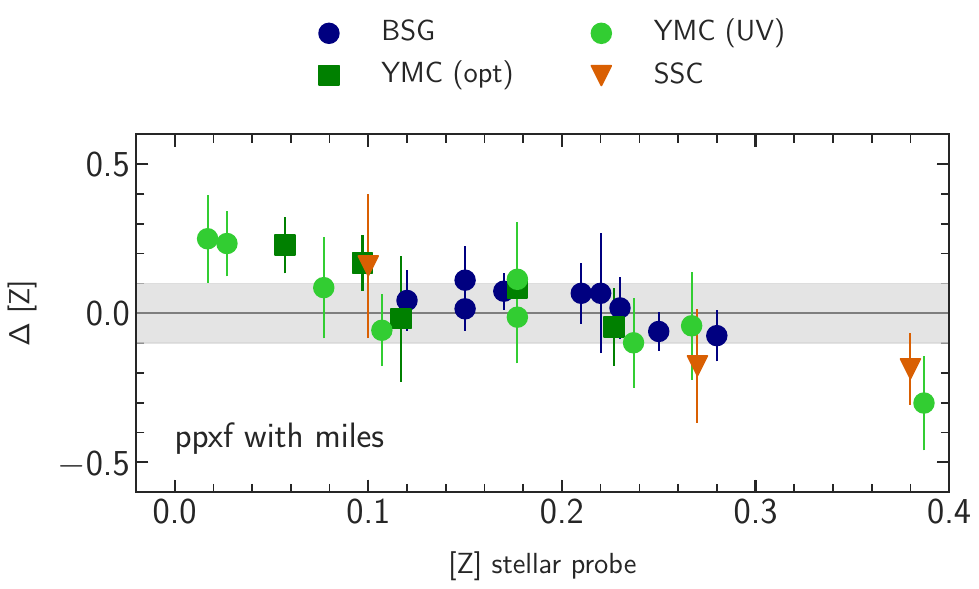}\medskip
    \center \includegraphics[width=0.95\columnwidth]{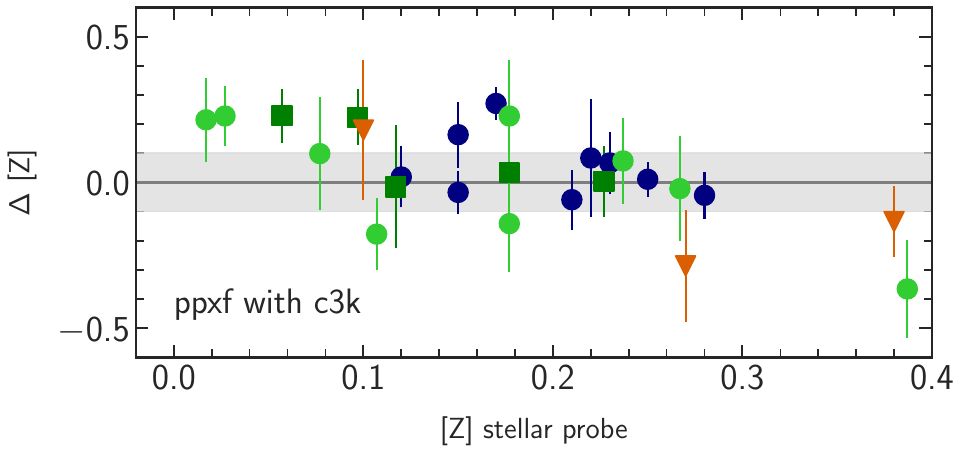}\medskip
    \center \includegraphics[width=0.95\columnwidth]{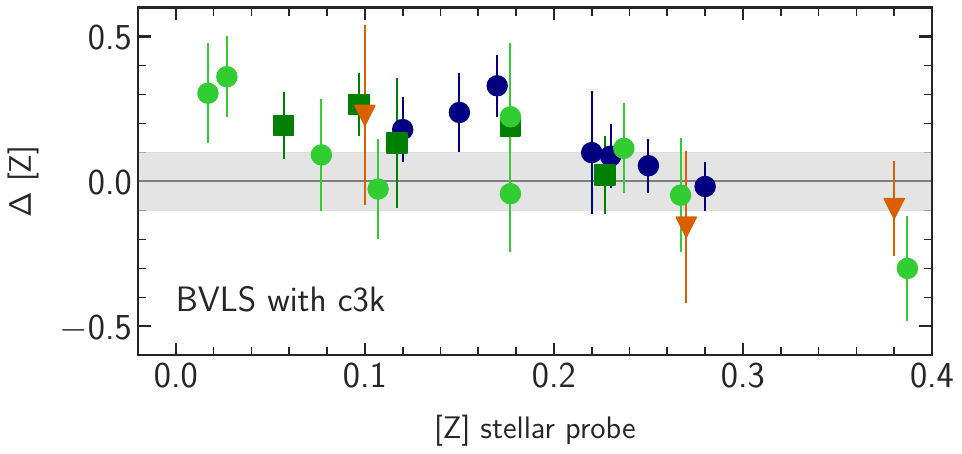}\medskip
	\caption{Difference $\Delta$[Z]$_y$ $-$ [Z]$_{probe}$ for different population synthesis fit methods and isochrone SSP. Top: pPXF with MILES. Middle: pPXF with C3K. Bottom: BVLS with C3K. Color coding is identical to Figure \ref{fig:stellar_probes} }    \label{fig:pPXFprobe}
\end{figure}
% - - - - - - - - - - - - - - - - - - - - - - - - - - - - - - - - - - - - - - - - - 

\subsection{Metallicity Comparison: Individual Stellar Sources vs. Population Synthesis}\label{subsec:compare}

As already mentioned in the introduction, M83 has been subject to detailed quantitative studies of individual luminous stellar sources: BSGs \citep{Bresolin2016}, SSCs \citep{Davies2017}, and YMCs \citep{Hernandez2018,Hernandez2019,Hernandez2021} (see Section 1). This provides us with the unique opportunity to compare their metallicities with our population synthesis results for the young stellar population. For all stellar sources within our field of view and in pixels not masked out, we carry out a one-to-one comparison of stellar source metallicity [Z]$_{probe}$ and population synthesis analysis metallicity [Z]$_y$ in the corresponding spatial bin. We exclude bins where the fraction of light coming from the young stellar population is less than 10\% and where the results are, therefore, quite uncertain.

Figure \ref{fig:stellar_probes} shows the detailed comparison. We find a very good agreement. The standard deviation of the difference $\Delta$Z = Z$_y$ $-$ Z$_{probe}$ is 0.12 dex and the mean value is 0.02 dex. There is no trend with the age of the stellar population and the light fraction contributed by young stars to the integrated spectra in each spatial bin. There is a weak indication of a small trend with Z$_y$, but that hinges mainly on one YMC point at low metallicity. In summary, this comparison demonstrates that the population synthesis approach is an accurate and powerful spectroscopic tool. The consistency between both approaches, spectroscopy of individual stellar probes and the analysis of the spectra of integrated stellar populations, is definitely reassuring.

\section{A comparison between different population synthesis approaches}\label{sec:approaches}

In the former sections of this work we presented the results obtained with one fitting code (a relatively simple $\chi^2$ minimization technique) and one set of stellar templates. It is interesting and important to repeat the work using alternative approaches. In our discussion we concentrate on metallicity, which according to our experience is more affected by the fitting algorithm than dust properties and ages. We will focus on the metallicity of the young stellar population [Z]$_y$, because this has been the center of our study in the previous sections. This is complementary to previous work, which analyzed old populations in globular clusters (see, for example, \citealt{Boecker2020, Geraldo2020}).

\begin{figure*}[htb!]
\centering
  \begin{minipage}[t]{0.48\textwidth}
    \includegraphics[width=\textwidth]{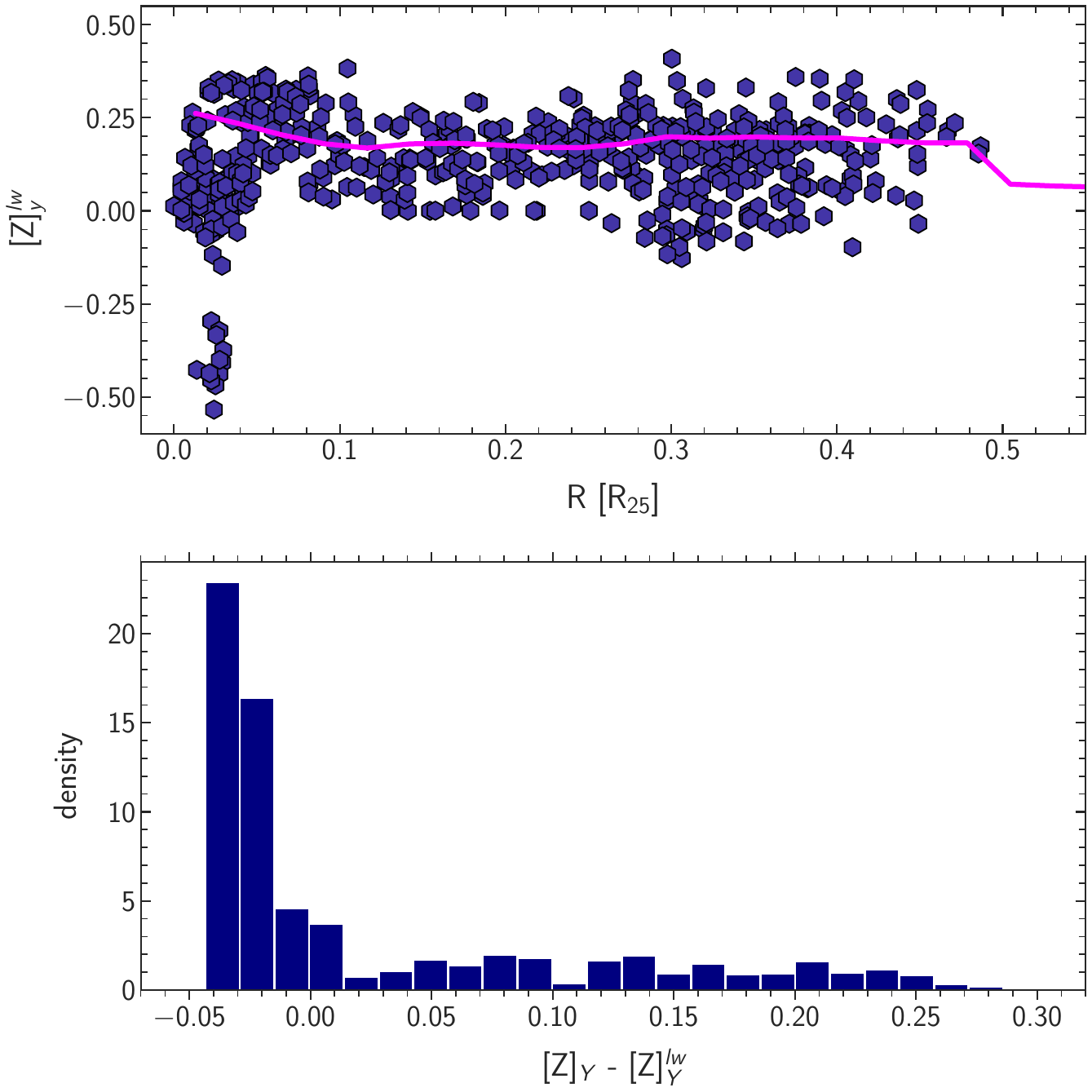}
  \end{minipage}
  \hfill
  \begin{minipage}[t]{0.46\textwidth}
    \includegraphics[width=\textwidth]{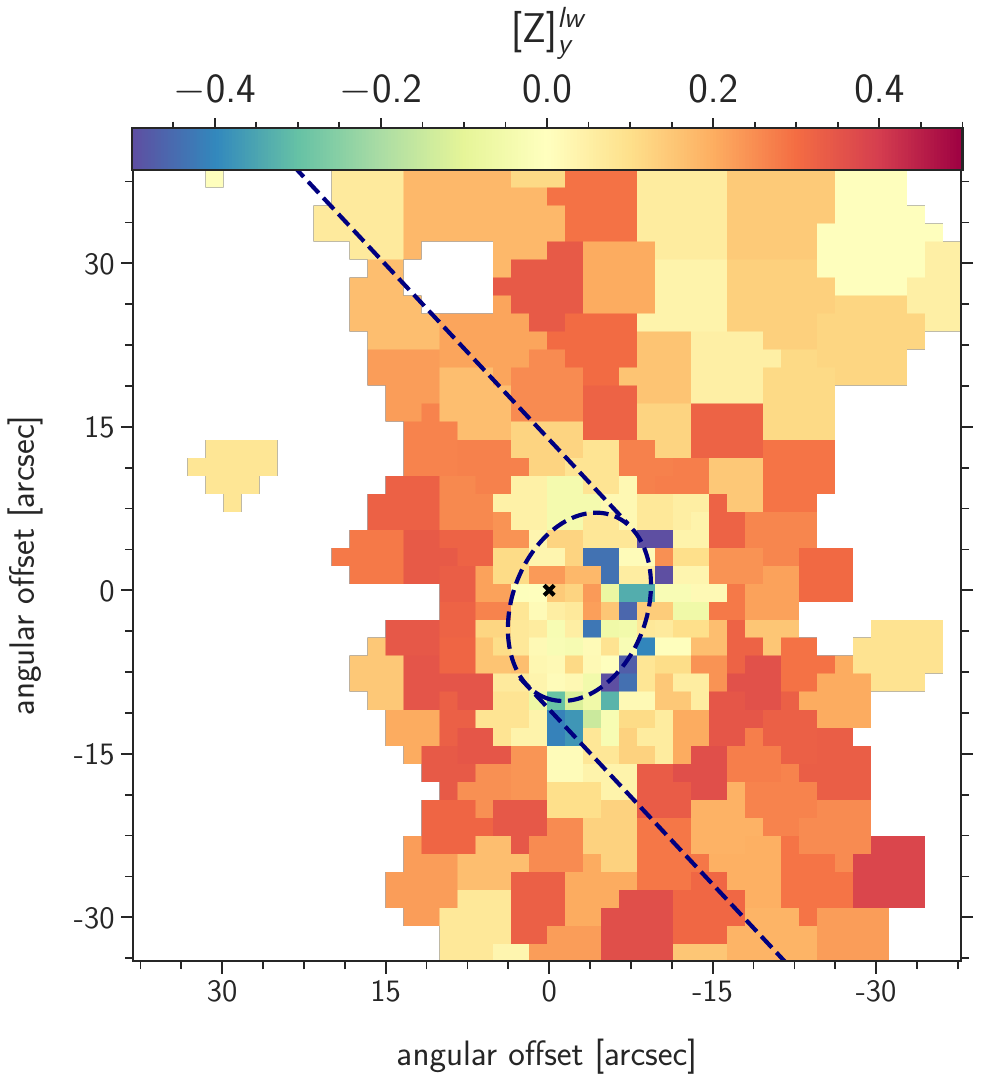}
  \end{minipage}
  \caption{Luminosity-weighted metallicity averages. Top left: Radial gradient of [Z]$_y^{lw}$, the chemical evolution model from \citet{Bresolin2016} is shown again in pink to guide the eye. Bottom left: Distribution function of the difference [Z]$_y$ $-$ [Z]$_y^{lw}$. Right: Central map of [Z]$_y^{lw}$ with the same color coding as in Figures \ref{fig:zmaplarge} and \ref{fig:pPXFfit}. }   \label{fig:lwfit}
\end{figure*}
% - - - - - - - - - - - - - - - - - - - - - - - - - - - - - - - - - - - - - - - - -  

In addition to our BVLS method ('bounded variable least-squares', see section \ref{sec:obs}), we also employ the pPXF (penalized PiXel-Fitting) fitting code \citep{Cappellari2023, Cappellari2004} with the Voronoi binned spectra to test the robustness of our approach. The base templates for the fit remain the same as before, i.e. the stellar library, isochrone and IMF are identical to BVLS. We also mask out the same emission line regions and use the same wavelength regime. The spectral analysis procedure, inspired by pPXF's GitHub examples\footnote{\url{https://github.com/micappe/ppxf_examples}}, follows a multi-step approach. We begin with outlier clipping, as described in \citet{Cappellari2023} Eq. 34, to remove spectral artifacts that could bias the fitting process. This step identifies and removes flux points that deviate by more than 3\,$\sigma$ in relative error. An multiplicative polynomial of degree 4 is switched on here allowing for slight continuum correction. Subsequently, a kinematic fit is performed to determine the radial velocity and velocity dispersion of the stellar population. This fit utilizes an additive polynomial of degree 4. The analysis progresses to the stellar population parameters. In this phase, the reddening is also determined, kinematic parameters are kept fixed from before, and the polynomial corrections are disabled to avoid degeneracies. This stage constitutes the primary fit for stellar ages, metallicities, and color excess. Finally, 'wild' bootstrapping is performed. The residuals from the last fit flip their sign at random and are then added to the obtained model fit. pPXF is then applied to each of these resampled models, generating distributions of stellar population parameters after 50 steps. This process enables robust uncertainty estimation and helps to identify potential biases or degeneracies in the fit, especially in central bins with a steep intrinsic SED, large extinction, and weak metal lines.

Figure \ref{fig:pPXFfit} summarizes the results obtained with pPXF as an alternative to our BVLS method. The radial metallicity distribution is very similar to that shown in Figure \ref{fig:zgradyoung}, except that the values are slightly larger ($\sim$ 0.05 dex). The distribution function of [Z]$_y^{pPXF}$ $-$ [Z]$_y^{BVLS}$ obtained from a spatial bin-by-bin comparison confirms the slight shift. However, most importantly, the standard deviation $\sigma$ = 0.09 dex is small and comparable to the uncertainties of the metallicity determination.  The central map compared to Figure \ref{fig:zmaplarge} confirms the low metallicites along the southern x$_2$ orbit, but also shows a few more low metallicity bins in the center inside the x$_2$ orbit. Although this is a slight quantitative difference, it generally confirms the existence of young metal-poor stars in the center.

Figure \ref{fig:pPXFprobe} shows the comparison with the metallicities of the stellar probes in a similar way to that carried out with our BVLS method in the previous section. The result is equally good. We find a standard deviation of $\Delta$[Z] = [Z]$_y^{pPXF}$ $-$ [Z]$_{probe}$ of 0.13 and a mean value of 0.02. This confirms the reliability of the population synthesis approach regardless of the fit method applied. We note, however, that the small trend with metallicity seems to be more pronounced than in our use of BVLS in Figure \ref{fig:stellar_probes}.

As described in Paper I, our isochrone SSPs are calculated from the Flexible Stellar Population Synthesis package (FSPS; version 3.2) \citep{Conroy2009, Conroy2010} using the MILES empirical library of stellar spectra \citep{Sanchez2006} augmented with a comprehensive set of spectra for young massive stars. An alternative option within FSPS is the C3K library \citep{Conroy2019,Byrne2022} based on theoretical model atmosphere spectra. In order to investigate the impact of the choice of SSPs we have carried a population synthesis analysis using spectra from C3K and the same MIST ioschrone set. This means that the metallicity and age grids remain the same as before. The resulting comparison with metallicities of the stellar probes is also shown in Figure \ref{fig:pPXFprobe}. Compared with the SSP based on the MILES library, the scatter is larger. We obtain $\sigma$ = 0.16 for both the pPXF and BVLS fitting methods. The mean value of $\Delta$[Z] is 0.03 dex for pPXF and 0.10 dex for BVLS, respectively. This means that the choice of the SSP has an influence. 

\section{Luminosity-weighted metallicity averages} \label{sec:lightweightmetal}

In galactic chemical evolution, stellar metallicity is defined as the ratio of the mass of metals confined in stars to the total stellar mass. Eq. \ref{eq:Zlog} takes this into account. However, many times and also in our previous work \citep{Sextl2023, Sextl2024} an alternative, simpler average is used:

\begin{align}
    [Z]^{lw} &= \sum_i b_i [Z]_i \\
    [Z]_y^{lw} &= \sum_{i_y} b_i [Z]_i.
    \label{eq:def_lw}
\end{align}

As a consequence of Eq.~\ref{eq:biNi}, this is a $V$-band luminosity-weighted average of the logarithmic metallicity. Although straightforward and intuitive, this definition of metallicity bears the risk that a relatively small number of stars with high luminosity can bias the average. The luminosities L$_i$ of the SSP isochrones depend on age with a strong and narrow peak between 5 and 8 Myr (see Figures 3 and 4 in \citealt{Sextl2023}). This peak can influence the luminosity-weighted average, as we will show in the following. 

Figure \ref{fig:lwfit} shows the luminosity-weighted metallicity of the young population [Z]$_y^{lw}$ obtained with our BVLS method. The radial distribution looks very similar to Figure \ref{fig:zgradyoung}, however, we notice subtle differences. In the central region and around the position of the spiral arms (0.3 to 0.4 R/R$_{25}$) we find shifts to lower metallicity. The distribution function of [Z]$_y^{BVLS}$ $-$ [Z]$_y^{lw}$ confirms this impression. The maximum of the distribution is very close to zero, but there is a tail extending to values as large as 0.3 dex, indicating that there are bins where the luminosity-weighted averages are significantly lower. Most of these bins are located in the center, as the central map of [Z]$_y^{lw}$ shows. Compared with Figure \ref{fig:zmaplarge} we now find many more central bins with a lower metallicity. The reason for this result is the presence of young stars with ages between 5 and 8 Myrs with lower metallicity, which contribute more to the luminosity-weighted average of the logarithmic metallicity because of their luminosity. In the linear average of metallicity masses, as given by Eq.~\ref{eq:Zlog} and Eq.~\ref{equation_Z}, these objects do not contribute much. Thus, while Eq.~\ref{eq:Zlog} describes the correct average in terms of chemical evolution, the luminosity weight in Eq.~\ref{eq:def_lw} indicates the presence of stars deviating from a homogeneous chemical evolution picture, confirming the presence of a low metallicity inflow. \\

\begin{figure}[htb!]
\centering
\medskip
       \includegraphics[width=\columnwidth]{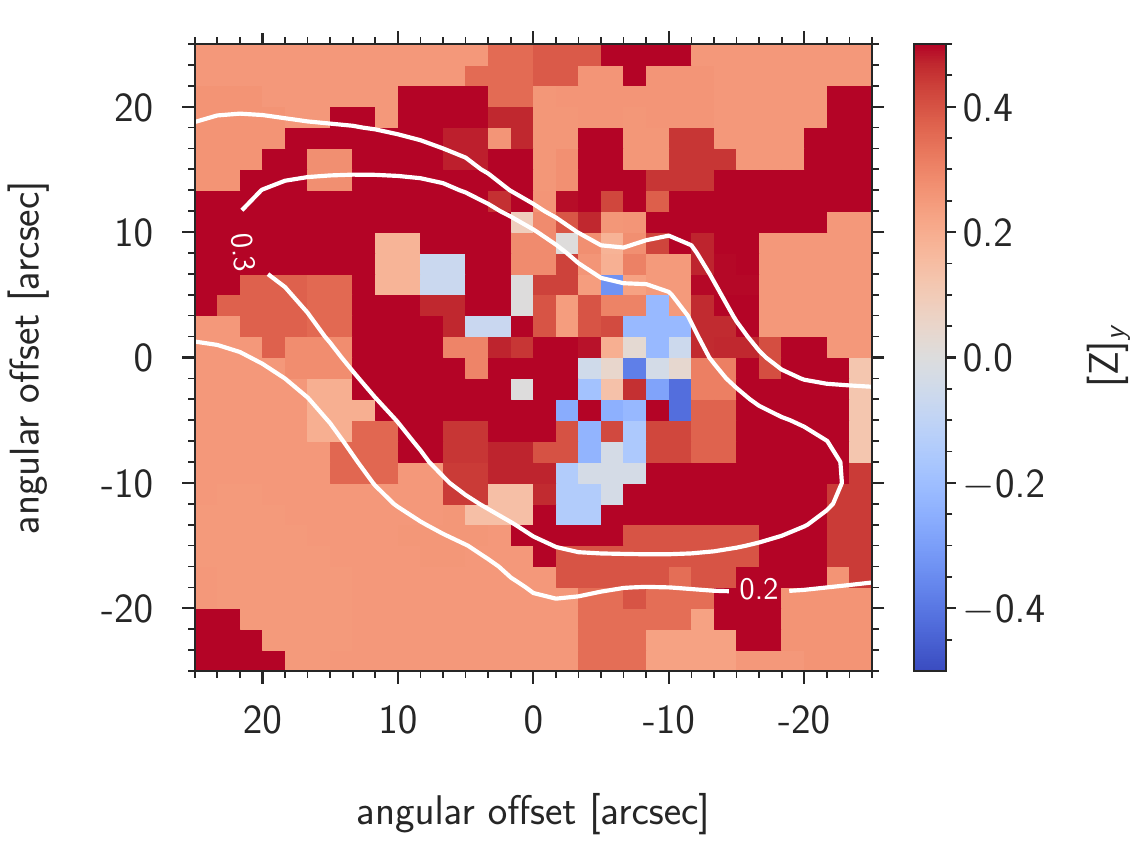}
  \caption{Reanalysis of NGC 1365: central map of [Z]$_y$}.   
 \label{fig:NGC1365}
\end{figure}

% - - - - - - - - - - - - - - - - - - - - - - - - - - - - - - - - - - - - - - - - - 

In our investigation in Paper I of the Great Barred Spiral NGC 1365 we have encountered a confined central region where the metallicity of the young population dropped dramatically and becomes lower than that of the old population. However, this result was based on the application of luminosity-weighted averages. We therefore have repeated the analysis of NGC 1365 calculating the metallicities [Z]$_y$ with the use of equation \ref{eq:Zlog} and assuming t$_{lim}^y$ = 1.6 Gyr as in Paper I. The result is shown in Figure \ref{fig:NGC1365}. Although there are quantitative differences from the luminosity-weighted results, the presence of metal-poor young stars in the center is confirmed. In the outer region of NGC1365 around $10$ kpc [Z]$_y$ remains the same but the gradient changes from $-0.020 \pm 0.003$ to $-0.010 \pm 0.003$ dex kpc$^{-1}$. 

% - - - - - - - - - - - - - - - - - - - - - - - - - - - - - - - - - - - - - - - - - 
\begin{figure*}[htb!]
       \medskip
	\center \includegraphics[width=1\textwidth]{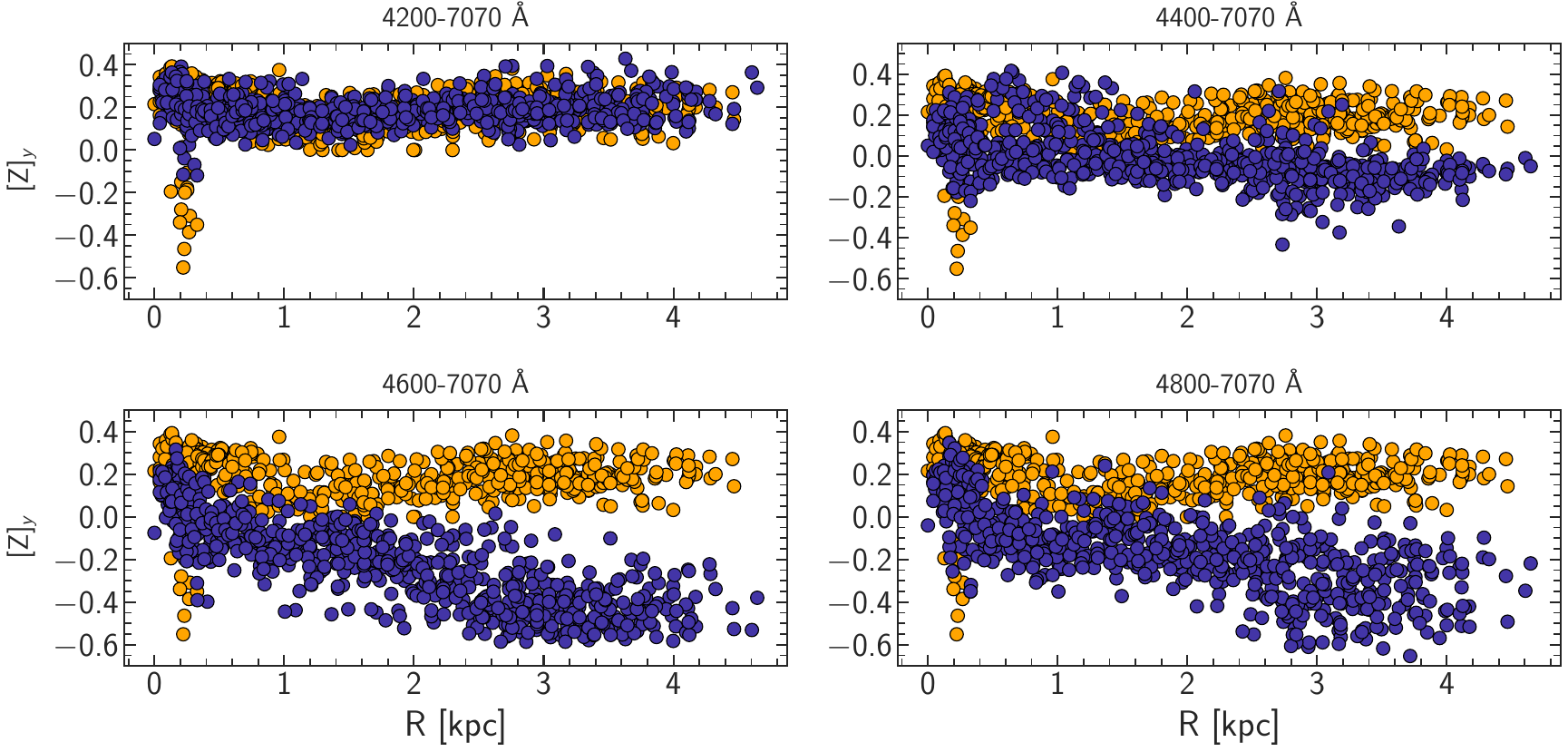}\medskip
	\caption{Effect of wavelength coverage on the metallicity of the young stellar population (age\,$<$\,100 Myr). The result from the baseline fit is drawn in each subplot in orange, the result with the wavelength cut in blue. The fitted wavelength is written over the fit each time and the figure margins were kept fixed for comparison.}   \label{fig:wavelength_z_y_new}
\end{figure*}

\section{The importance of blue wavelength coverage} \label{sec:wcoverage}
For our full spectral fitting analysis, we utilize the spectral range of 4000 to 7070 \AA, where the TYPHOON spectra have the best signal. As we have already pointed out in Paper I, for the characterization of the young stellar population, this is an important blueward extension compared to the range 4800 to 7000 \AA\ used in corresponding studies of other IFU surveys such as PHANGS-MUSE (see, for instance, \citealt{Pessa2023}). For young stars, the blue part of the optical regime is crucial. In addition to the steepening SED it contains numerous spectral features sensitive to hot, massive stars \citep{Vazdekis2016}. These features are essential for accurately determining the ages and metallicities of young stellar clusters and recent star formation episodes. In this section, we demonstrate the crucial importance of blue wavelength coverage. We investigate what happens with the characteristics of the young stellar populations, most importantly their metallicity, when parts of the blue optical spectrum are removed in the fit. Is it still possible to successfully characterize the young population with a wavelength coverage starting at 4800 $\angstrom$? 

For this purpose, we fit all spatial Voronoii bins again with the same SSP templates, but with varying wavelength ranges of [4200\,$\angstrom$, 7000\,$\angstrom$], [4400\,$\angstrom$, 7000\,$\angstrom$], [4600\,$\angstrom$, 7000\,$\angstrom$] and [4800\,$\angstrom$, 7000\,$\angstrom$]. Figure \ref{fig:wavelength_z_y_new} summarizes the results. In each plot, the metallicity of the young stellar population in each bin is displayed against the baseline fit.

The plots in the top row exhibit only minor deviations, though these fits already miss prominent spectral features such as He I at 4026\,$\angstrom$ and the Fraunhofer e line. However, in the second row with blue wavelengths starting at 4600\,\AA\ and 4800\,\AA, respectively, we start to see significant effects. A metallicity gradient is obtained, which is not real. This is because the reduced wavelength range is less efficient in breaking the age-metallicity degeneracy and begins to produce erroneous metallicities depending on the light fractions and ages of the young population.

%The plots in the top row exhibit only minor deviations, though these fits already miss prominent spectral features such as He I at 4026$\angstrom$ and the Fraunhofer e and G lines. The reduced wavelength range complicates breaking the age-metallicity degeneracy. A closer look reveals that the fraction of young stars is now higher than in our standard case which leads as a consequence to lower mean flux-weighted ages (the slightly larger log age$_y$ gets cannot outweigh this) as well as lower metallicities. That effect then pushes through to the young stellar component. Starting at the second row (4600$\angstrom$ and upwards) it is not reasonable anymore to recover a proper young stellar population at this low spectral-resolution. The loss of metal-sensitive features and spectral indicators for different age groups leads to larger error bars on derived parameters and a reduced confidence in the population split. A negative gradient stars to appear and the algorithm detects increased light-fraction of very young populations ($<$ 10Myr) (figure \ref{fig:wavelength_by}). Also, the larger the light fraction of young stars is to begin with, the better is the overall situation as the remaining optical spectrum contains enough information about them. \\
This effect is seen in a similar manner in the mock experiments by \citet{Lee2023} with age-divided populations. The authors interpret their findings as a result of a degeneracy not often discussed: a stellar population composed mostly of old stars with a tiny fraction of very young ones can produce a spectrum similar to one with more moderately young stars but still dominated by older stars. The bluer part of the spectrum is crucial to mitigate this.  \\
As a further step, we ran calculations cutting off the blue part of the optical regime as before, but we also extended the wavelength regime towards the red part of the spectrum in order to have in total a 3000\,$\angstrom$ range available for the fit. This does not change the situation and leads to identical trends. \\ 

The overall conclusion we derive is that age-divided mean stellar populations are sensitive towards the wavelength coverage of the fitted spectra - at least at the relatively low spectral resolution at which the TYPHOON survey operates. To what extent a spectral resolution higher than that reached by the TYPHOON survey ($R=800$) can improve the metallicity determination of the young stellar population cannot be answered here. We will investigate this in a follow-up publication.\\

\section{Summary} \label{sec:discussion}
We use a population synthesis full spectral fitting method of TYPHOON 3D spectral data to investigate the inner 4~kpc of the disk of the nearby face-on barred spiral galaxy M83. Since M83 is characterized by intense star formation activity with ongoing star formation bursts, we focus on the young stellar population and the distribution and properties of interstellar dust and its correlation with molecular gas.

The average reddening produced by dust is about 0.2 mag, but we find a strong concentration of dust in the star-forming regions related to the spiral arms, along the bar, and in the center. Close to the center, we detect a dust cavity with a 260 pc diameter similar in size to the local bubble in the Milky Way. The distribution of the ISM molecular gas traced by CO(2-1) emission is correlated with the distribution of dust, supporting the scenario in which molecular gas forms on the surface of dust grains.

We find a wide range of R$_V$ distributed throughout the galactic disk. In the center, we detect an interesting anticorrelation between R$_V$ and the relative column densities of the ISM molecules CCH, CN, and CS. The presence of these molecules is usually attributed to ISM photo-dissociation regions with dust grains of smaller sizes. Our result confirms this.  

The population synthesis also allows us to estimate stellar ages. Stars in the central region with strong star formation are very young, with ages of about 5 Myr. The stars in the dust cavity are slightly older with 7 Myr (consistent with \citet{Harris2001} and \citet{Jones2024}). Outside the central region, we determine the time when 85 percent of the mass of the total stellar population has formed. We find a weak gradient, indicating rapid inside-out growth of the stellar disc.

The young stellar population in the inner 4 kpc of the disk of M83 has a supersolar metallicity [Z]$_y$ = 0.2 dex. The radial distribution is flat, without any indication of a negative gradient, and agrees well with the simple chemical evolution model presented in \citet{Bresolin2016}. The metallicity of the old population is 0.4 dex lower. In the central region, we find a significant drop of [Z]$_y$. We attribute this to infall of metal poor gas from the circumgalactic medium or a dwarf galaxy interloper that affects subsequent star formation. An alternative is AGN-interrupted chemical evolution as discussed in Paper I.

We check our full spectral fitting method by the comparison with metallicities obtained through the detailed spectroscopic analysis of individual BSGs, SSCs and YMCs. We find good agreement without indications of a systematic effect. To our knowledge, this is the first test of this kind confirming the reliability of the method. In order to further assess the uncertainties we also employed an alternative full spectral fitting algorithm (pPFX) as well as an alternative set of single stellar population spectra (C3K), and repeated the analysis. The results were basically confirmed. However, we obtained a small trend with metallicity and a larger scatter with C3K.

In our analysis, we measure the mass of metals confined in the stellar population, which is then used to determine the metallicity. Our research indicates that using a simple luminosity-weighted average of logarithmic metallicity across stellar populations of varying ages can be misleading. This method is particularly problematic when the stellar mix includes a small but bright group of young, metal-poor stars. In such cases, the results may not accurately represent the true metallicity distribution of the entire stellar population.

The analysis of the young stellar population as described above requires a wide spectral coverage of the IFU spectra obtained, most importantly with a far extension to blue wavelengths. We show that restrictions of the optical wavelength range on the blue side have significant effects and introduce large uncertainties.

Our findings have significant implications for studies of galactic chemical evolution and validate the use of integrated light spectroscopy for metallicity determinations in unresolved stellar populations.

\begin{acknowledgments}
Acknowledgments. This work was initiated and supported by the Munich Excellence Cluster Origins funded by the Deutsche Forschungsgemeinschaft (DFG, German Research Foundation) under Germany's Excellence Strategy EXC-2094 390783311. ES acknowledges support for the COMPLEX project from the European Research Council (ERC) under the European Union's Horizon 2020 research and innovation
program grant agreement ERC-2019-AdG 882679. KG is supported by the Australian Research Council through the Discovery Early Career Researcher Award (DECRA) Fellowship (project number DE220100766) funded by the Australian Government. We also deeply acknowledge the stimulating atmosphere and scientific discussion at a four week workshop program organized by the Munich Institute for Astro-, Particle-, and BioPhysics (MIAPbP). MIAPbP is also funded by the Deutsche Forschungsgemeinschaft (DFG, German Research Foundation) under Germany's Excellence Strategy EXC-2094 390783311. We thank our colleagues Kathryn Kreckel, Andreas Burkert, Giovanni Tedeschi and Til Birnstiel for their inspiring interest in our work. Finally, we wish thank our referee for a very constructive and detailed critical report.
\end{acknowledgments}

%\vspace{5mm}
%\facilities{CTIO:1.5m,du Pont 2.5-m telescope (+ WFCCD) at Las Campanas Observatory}

%\software{Astropy \citep{astropy2013,astropy2018,astropy2022}  }

\bibliography{M83_TYPHOON}{}
\bibliographystyle{aasjournal}

\end{CJK*}
\end{document}